\documentclass{article}

\usepackage{booktabs}
\usepackage{graphicx}
\usepackage{amssymb}
\usepackage{cite}

\newcommand{\radian}{\mbox{rad}}
\newcommand{\joule}{\mbox{J}}
\newcommand{\megahertz}{\mbox{MHz}}
\newcommand{\BoltzmannConstant}{\ensuremath{k_{\mbox{\scriptsize B}}}}
\newcommand{\Kelvin}{\mbox{K}}
\newcommand{\Edissipated}{\ensuremath{E_{\mbox{\scriptsize dissipated}}}}
\newcommand{\Kdrag}{\ensuremath{k_{\mbox{\scriptsize rd}}}}
\newcommand{\second}{\mbox{s}}
\newcommand{\nanometer}{\mbox{nm}}
\newcommand{\kilogram}{\mbox{kg}}
\newcommand{\meter}{\mbox{m}}
\newcommand{\piconewton}{\mbox{pN}}
\newcommand{\nanonewton}{\mbox{nN}}

\begin{document}

\title{Mechanical Computing Systems Using Only Links and Rotary Joints}
\author{RALPH C. MERKLE \and ROBERT A. FREITAS JR. \and TAD HOGG \and THOMAS E. MOORE \and MATTHEW S. MOSES \and JAMES RYLEY}

\maketitle  

\begin{abstract}
A new model for mechanical computing is demonstrated that requires only two basic parts: links and rotary joints. These basic parts are combined into two main higher level structures: locks and balances, which suffice to create all necessary combinatorial and sequential logic required for a Turing-complete computational system. While working systems have yet to be implemented using this new approach, the mechanical simplicity of the systems described may lend themselves better to, e.g., microfabrication, than previous mechanical computing designs. Additionally, simulations indicate that if molecular-scale implementations could be realized, they would be far more energy-efficient than conventional electronic computers.
\end{abstract}

\section{Introduction}
Methods for mechanical computation are well-known. Simple examples include function generators and other devices which are not capable of general purpose (Turing-complete) computing (for review, see \cite{Reif_2009}), while the earliest example of a design for a mechanical general purpose computer is probably Babbage's Analytical Engine, described in 1837 \cite{Bromley_1983}. One of the very first modern digital computers was a purely mechanical device: the Zuse Z1, completed in 1938 \cite{Zuse_1993}. 

At a time when silicon-based electronic computers are pervasive, powerful, and inexpensive, the motivation for studying mechanical computer architectures is not immediately obvious. However, many research groups are currently investigating mechanical, electromechanical, and biochemical alternatives to conventional semiconductor computer architectures because of their unique potential advantages. For example, mechanical systems can withstand much higher temperature and radiation exposure than their electronic counterparts, and hence may be useful in certain niche applications \cite{Bradley_2003,Pott_2010,Kam_2011,Chowdhury_2013}. DNA computing makes use of vast numbers of molecules to solve computational problems in parallel \cite{Boruah_2015,Okamoto_2004}. 

One potential advantage of these many alternative computing architectures is energy efficiency. The new mechanical approach presented in this paper is particularly well-suited for implementing physically reversible logic gates. Reversible logic gates are one alternative technology that can, in principle, sidestep fundamental limitations of complementary metal-oxide-semiconductor (CMOS) transistors, and thus facilitate computers that operate with vastly reduced energy dissipation \cite{Athas_1994,Moon_1996,Roukes_2004,Wenzler_2014,Frank_2017}.  

While previous designs for mechanical computing vary greatly, the few designs capable of general purpose computing require a substantial number of basic parts, such as various types of sliding plates, gears, linear motion shafts and bearings, springs (or other energy-storing means), detents, ratchets and pawls, and clutches.

The use of many parts brings with it a number of potential problems, such as increased friction, higher mass, and increased device complexity. Such issues can reduce performance and increase the difficulty of manufacturing. However, reducing the number, complexity, and mass of parts in a mechanical computer is not a simple task due to the need to provide both universal combinatorial logic (e.g., AND, NAND, NOR, etc.) and sequential logic (memory).

Sequential logic in particular, being the basis for memory, requires the ability to conditionally decouple logic elements from current inputs. This is because memory cannot be only a deterministic result of just current inputs, otherwise previous states cannot be saved. Storing information, which is easily accomplished in electronic systems using latches or flip-flops, is not as easily accomplished in a mechanical system which may have to actually connect and disconnect parts of the system from each other (e.g., using a clutch-like mechanism) at appropriate times. This paper demonstrates that mechanical computers can be greatly simplified by using only two parts: links and rotary joints.

\begin{figure}
\centerline{\includegraphics[width=4in]{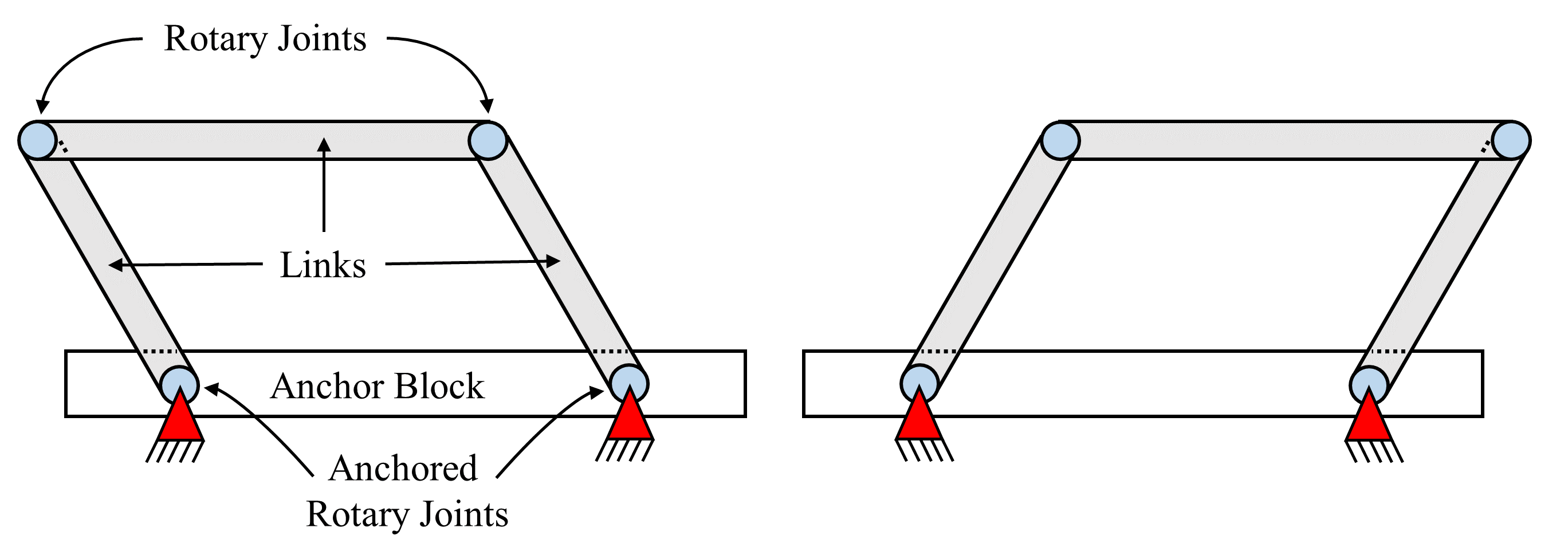}}
\caption{A 4-Bar Linkage in two configurations, left-leaning (left) and right-leaning (right)}
\label{fourbar-left-right-leaning}
\end{figure}

\begin{figure}
\centerline{\includegraphics[width=4in]{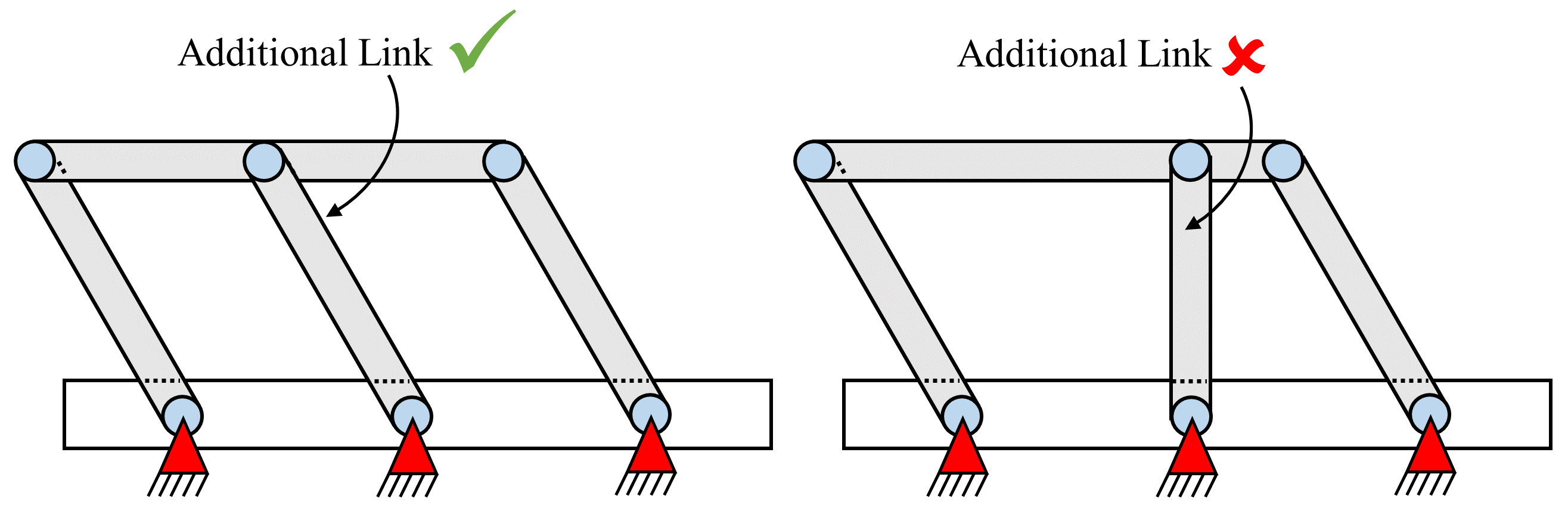}}
\caption{The mobile linkage (left) is free to move, while the non-mobile linkage (right) is static.}
\label{foubar-mobile-nonmobile}
\end{figure}

\section{Computing With Only Two Parts}
Links are stiff, truss-like structures. Rotary joints are used to connect links in a manner that only allows rotational movement in a single plane. Perhaps counterintuitively, no type of clutch-like mechanism is required. All parts of the system can remain permanently connected and yet still provide all necessary combinatorial and sequential logic.

\begin{figure}
\centerline{\includegraphics[width=4in]{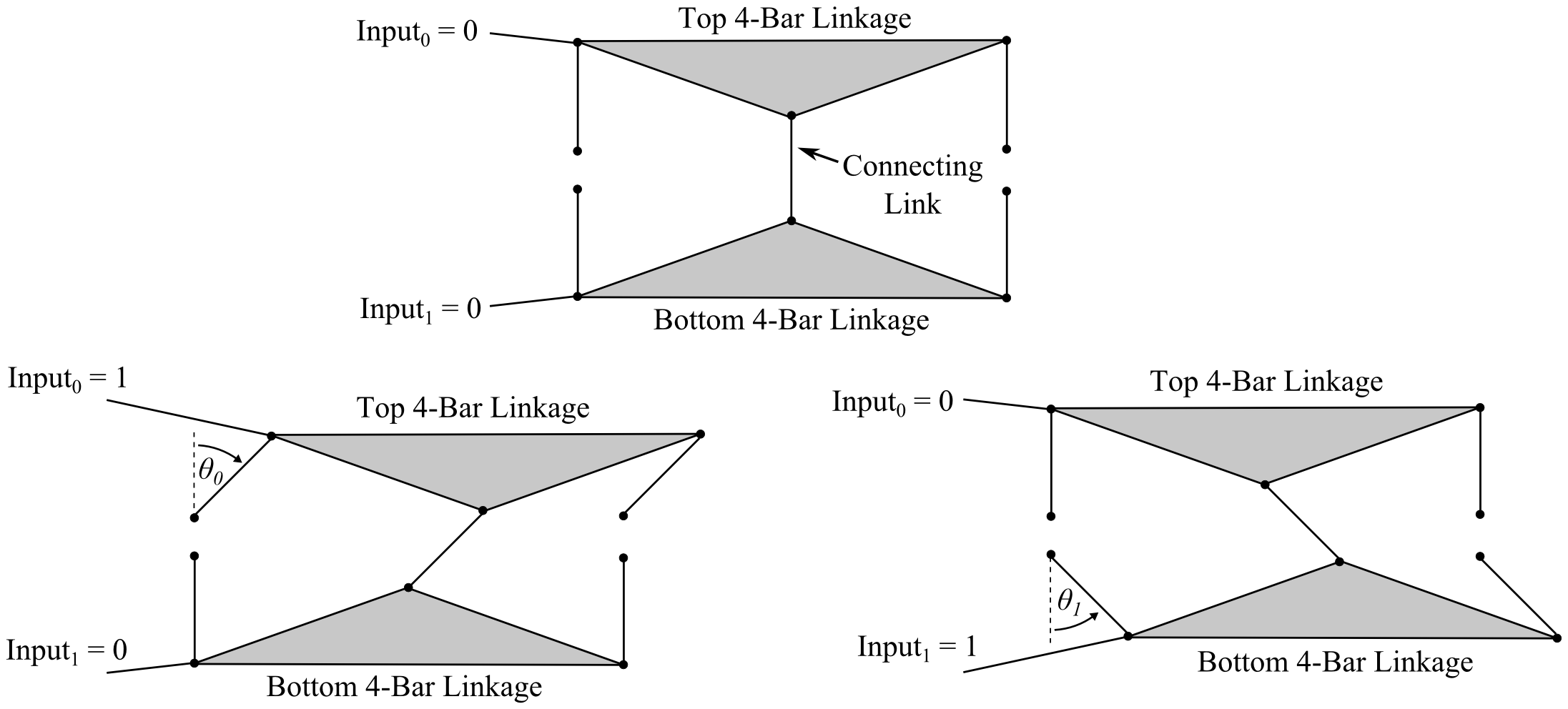}}
\caption{A lock in the (0, 0) position (top), a lock in the (1, 0) position (bottom left), and a lock in the (0,1) position (bottom right). The (1,1) position is prohibited by the linkage geometry. See Fig.~\ref{locking force} for additional discussion.}
\label{lock-diagram}
\end{figure}

\subsection{4-Bar Linkages}
To demonstrate how this can be accomplished, we start with a simple, well-known mechanism, the 4-bar linkage (also referred to as a 3-bar linkage in some literature -- the 4th bar is provided by the anchor block or base and is sometimes ignored for naming purposes). The 4-bar linkage relies only upon links and rotary joints. A 4-bar linkage can rotate around its anchored rotary joints (denoted by a circle with a triangle, while unanchored rotary joints are just circles), allowing, for example a ``left-leaning'' configuration to rotate into a ``right-leaning'' configuration (see Fig.~\ref{fourbar-left-right-leaning}).

Note that, while a 4-bar linkage could assume many positions, we focus on the use of two distinct positions. This is because two positions can be used to signify 0 and 1, which is convenient when creating a system for binary computing. The actual angle traversed by the links when moving from, e.g., left-leaning to right-leaning is not critical. As diagrammed, it is approximately 45 degrees, but could be more or less as long as the design supports reliably differentiating between two positions; one representing 0, and the other representing 1.

It is important to note that in a parallelogram-type (meaning, the two side links must be the same length) 4-bar linkage, if an additional link is added to the center of the linkage, as long as the length and angle is the same as the side links, the linkage will still rotate around the anchored rotary joints. However, if the additional link is not the same length, or is not at the same angle, as the other links, the mechanism will not move (see Fig.~\ref{foubar-mobile-nonmobile}). This is because the additional link will be attempting to pivot through a different arc than the side links. The effect of this is that the side links and the center link are trying to move the top link in two different directions at once. This results in the mechanism binding up, or ``locking.'' This locking behavior is important in the creation of ``locks,'' one of two main higher-level mechanisms used by the system we describe (the other being the ``balance'').

\subsection{The ``Lock"}
A “lock” is a mechanism composed of two 4-bar linkages, connected in the center via a connecting link (see Fig.~\ref{lock-diagram}). The connecting link is the same length as the two side links of each 4-bar linkage, and, in the starting position, is parallel to all four side links. 

In the implementation depicted, the connecting link is affixed to two extra links at the top, and two at the bottom, each pair of which form a rigid triangle with the rest of the respective 4-bar linkage. This triangular projection allows the connecting link to be the same length as the side links. If the triangular projection were not present and the connecting link had to connect directly to the horizontal links of the 4-bar linkages, it would not be the same length as the side links. As already mentioned, this would cause the 4-bar linkages to bind up, or lock. To function in the intended manner, all of the vertical links must be the same length and must, at least initially, be parallel to each other.

Obviously, linkages need not have their two positions be left-leaning and right-leaning. Rather, for binary computing purposes, they just need to have two distinct positions which can represent 0 and 1. Figure \ref{lock-diagram} uses the convention that an input of 0 results in the side links being vertical, while an input of 1 would cause them to lean to the right.

\begin{figure}
\centerline{\includegraphics[width=4in]{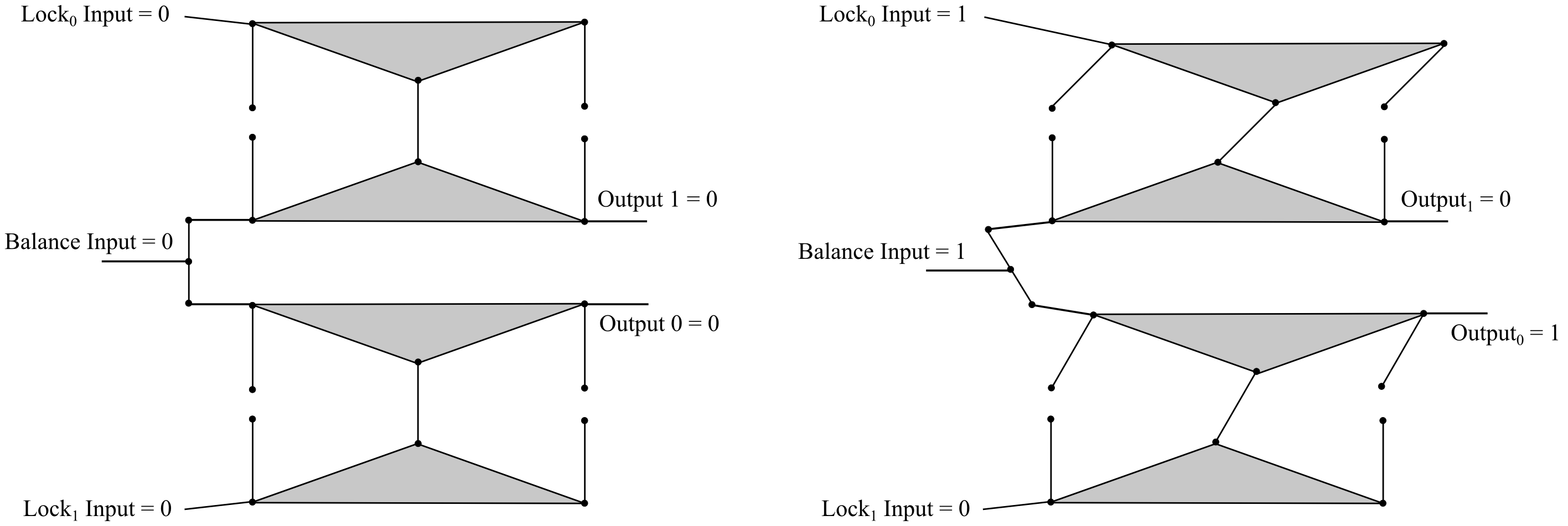}}
\caption{A balance coupled to two locks. The inactive configuration is shown on the left. On the right, $Lock_0 \, Input$ has been activated, followed by activation of the $Balance \, Input$, which in turn activates $Output_0$.}
\label{lock-and-balance}
\end{figure}

Representative inputs, in the form of linear slides $Input_0$ and $Input_1$, have been added to Fig.~\ref{lock-diagram}, to show the direction of actuation and where input could be connected. It is assumed that there is one input for each 4-bar linkage, or one input each for the top and bottom of the lock. In an actual system, lock inputs would be connected to other parts of the system (e.g., data from memory, or clock signals). Linear slides are not required.

The lock position depicted in the top of Fig.~\ref{lock-diagram} would be called the (0,0) position, referring to the state of the inputs. The lefthand portion of Fig.~\ref{lock-diagram} shows a depiction of the (1,0) state, where $Input_0$ has been set to 1 and thereby pushed its respective 4-bar linkage into the right-leaning configuration. The (0,1) state is shown in the righthand portion.

The (1,1) state is not possible, and this is a key aspect of a lock. Once one of the inputs has been set to 1, one side of the lock rotates around both its anchored rotary joints, and its connecting link. Crucially, while the connecting link never changes length, it does change angle. As depicted in Fig.~\ref{lock-diagram}, once the lock moves into the (1,0) position, the connecting link is still parallel to the side links of the top 4-bar linkage, but is no longer parallel to the side links of the bottom 4-bar linkage. Due to the requirement that, if a 3rd link is present, this link be both the same length and at the same angle as the side links, once either the top or bottom 4-bar linkage moves, the other cannot; it is locked. It is for this reason that the (1,1) position is impossible. Once the lock has moved from (0,0) to (1,0) or (0,1), the only possible movement is back to the (0,0) position. Once back to the (0,0) position, either input could be set to 1, but both inputs can never be set to 1 at the same time.

Note that flexures \cite{Howell_2001,Ion_2017} could take the place of rotary joints, allowing a lock (and other structures) to be monolithic, potentially simplifying manufacture using, for example, MEMS or NEMS techniques \cite{Ekinci_2005,Fettig_2001,Bradley_2003,Chowdhury_2013,Sharma_2009,Modi_2007,Pott_2010}. One might even argue that by using flexures, an entire computing system can be largely-monolithic, but as that raises the question of just what a part is, we will use links and rotary joints for clarity. Note that both flexures and rotary joints can be quite energy-efficient as they largely avoid sliding friction (which can have the added benefit of reducing wear \cite{Skakoon_2009}).

\subsection{The ``Balance"}
The balance, so named because it is superficially similar to a classic pan balance, connects an input to a link which has three rotary joints: one at each end, and one in the center. The input is connected to the center rotary joint. The two end rotary joints are connected to other structures. Most commonly, these other structures are locks. The result is that, when the balance input is changed, one of the side rotary joints remains stationary, while the other moves. Which side is stationary and which side moves can be determined by data inputs. How this works in practice is best illustrated by example. Figure \ref{lock-and-balance} depicts a balance connected to two locks. While the diagrammatic representations have been simplified slightly, it will be obvious that each lock corresponds to the lock mechanism depicted in Fig.~\ref{lock-diagram}.

The only difference between the locks in Fig.~\ref{lock-and-balance} and the lock in Fig.~\ref{lock-diagram} is that in Fig.~\ref{lock-and-balance}, rather than having two individual inputs, each lock has one input that is specific to a given lock, and the other input is supplied indirectly by the balance, which also has an input. Note that input to balances will often be supplied by a clock system. Like conventional computers, a multiphase clocking system is required for the described computing system. Such a clocking system is assumed to be present; describing how to implement such a system is beyond the scope of this document, although this too can be accomplished with only links and rotary joints, if desired.

There are assumed to be two main rules governing the mechanism shown in Fig.~\ref{lock-and-balance}, and these rules would be enforced by the overall system; they are not inherent in the mechanism depicted:
\begin{enumerate}
\item Either the $Lock_0 \, Input$, or the $Lock_1 \, Input$, but never both, must be set to 1.
\item The lock inputs and the balance input must be set sequentially, not simultaneously.
\end{enumerate}

Given these rules, the operation of the mechanism is as follows. Each step could be thought of as a different clock phase (which implicitly enforces the second rule above):
\begin{enumerate}
\item The mechanism starts with all inputs set to 0.
\begin{itemize}
  \item This means that neither lock is locked. 
\end{itemize}
\item Either the $Lock_0 \, Input$ is set to 1, or the $Lock_1 \, Input$ would be set to 1.
\begin{itemize}
  \item This results in locking one of the locks.
  \item Locking one of the locks constrains which side of the balance can move.
\end{itemize}
\item The balance's input is set to 1.
\begin{itemize}
  \item Since only one side of the balance is free to move, the balance input is transmitted down the path which is free to move.
  \item This produces an output of 1 at either $Output_0$ or $Output_1$, depending on which lock input was set to 1.
\end{itemize}
\end{enumerate}

This provides one example of performing simple logic that can, for example, be used to route data, shunting inputs down one path or another based on the states of $Lock_0$ and $Lock_1$. 

Note that this simple logic and conditional routing has been accomplished using only links and rotary joints, which are solidly connected at all times. No gears, clutches, switches, springs, or any other mechanisms are required (keeping in mind that the input mechanisms shown are representative; an actual system may use a different input implementation).

\begin{figure}
\centerline{\includegraphics[width=\columnwidth]{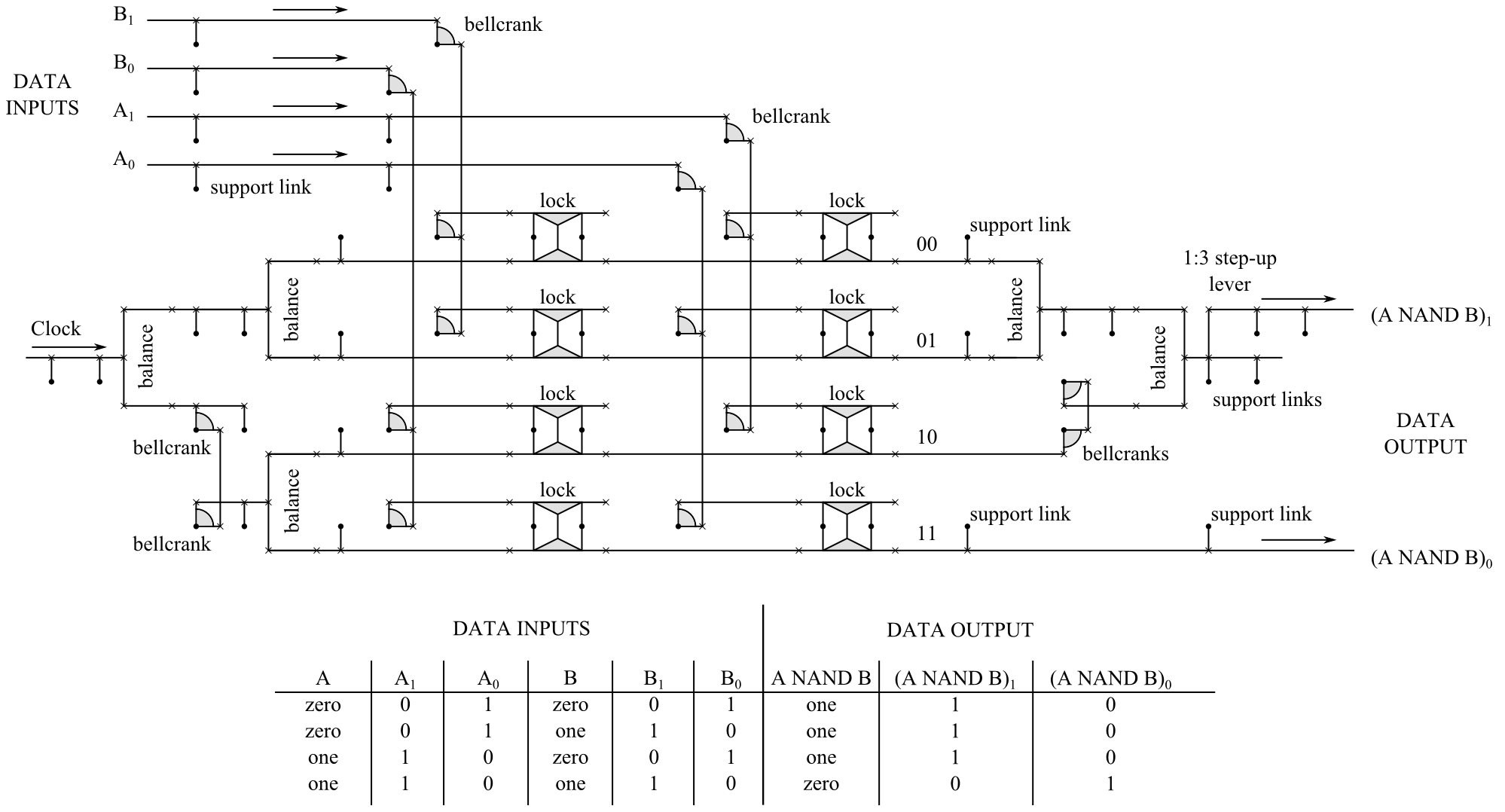}}
\caption{A balance- and lock-based NAND Gate, using dual rail logic (i.e. two-links per bit).}
\label{NAND-gate}
\end{figure}

\section{Universal Combinatorial Logic}
\label{UCL}
Although the previous example could be thought of as performing simple logic, it is perhaps more useful for routing data. The mechanism in Fig.~\ref{lock-and-balance} cannot provide all the logic necessary for a complete computational system. However, it is possible to create all necessary logic using nothing but locks and balances (and a few extra links and rotary joints to route and/or copy data). 

Any traditional 2-input logic gate, including AND, NAND, NOR, NOT, OR, XNOR and XOR, can be created directly from the appropriate combination of locks and balances. While most of these gates are illustrated in the Appendix, there is no need to address each in detail to prove that universal logic can be created using links and rotary joints. This is because it is well-known that NAND alone suffices to create all necessary combinatorial logic (i.e., all other logic can be created from combinations of NAND gates \cite{Feynman_1985}). Therefore, as a proof of the fact that links and rotary joints suffice to create the combinatorial logic required for a Turing-complete computing system, Fig.~\ref{NAND-gate} shows how a NAND gate can be implemented. (Note that reversible gates can also be created using only links and rotary joints, and a Fredkin gate is also demonstrated in the Appendix).

\begin{figure}
\centerline{\includegraphics[width=\columnwidth]{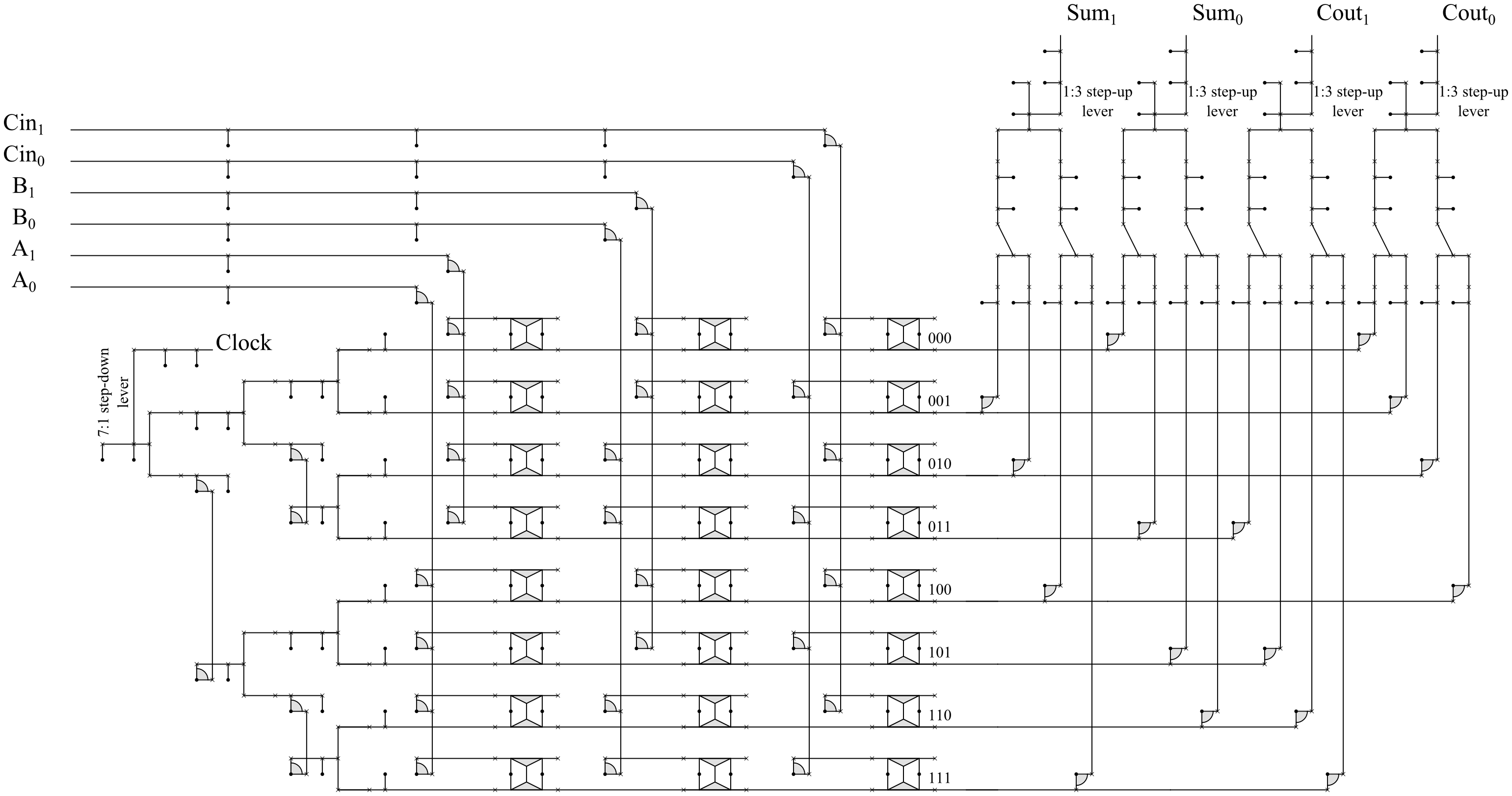}}
\caption{A 1-bit full adder, performing the logical operation described in Fig.~\ref{full adder table}. A table of the schematic symbols used in this drawing is provided in the Appendix (Fig.~\ref{schematic symbols}).}
\label{full adder}
\end{figure}

\begin{figure}
\centerline{\includegraphics[width=4in]{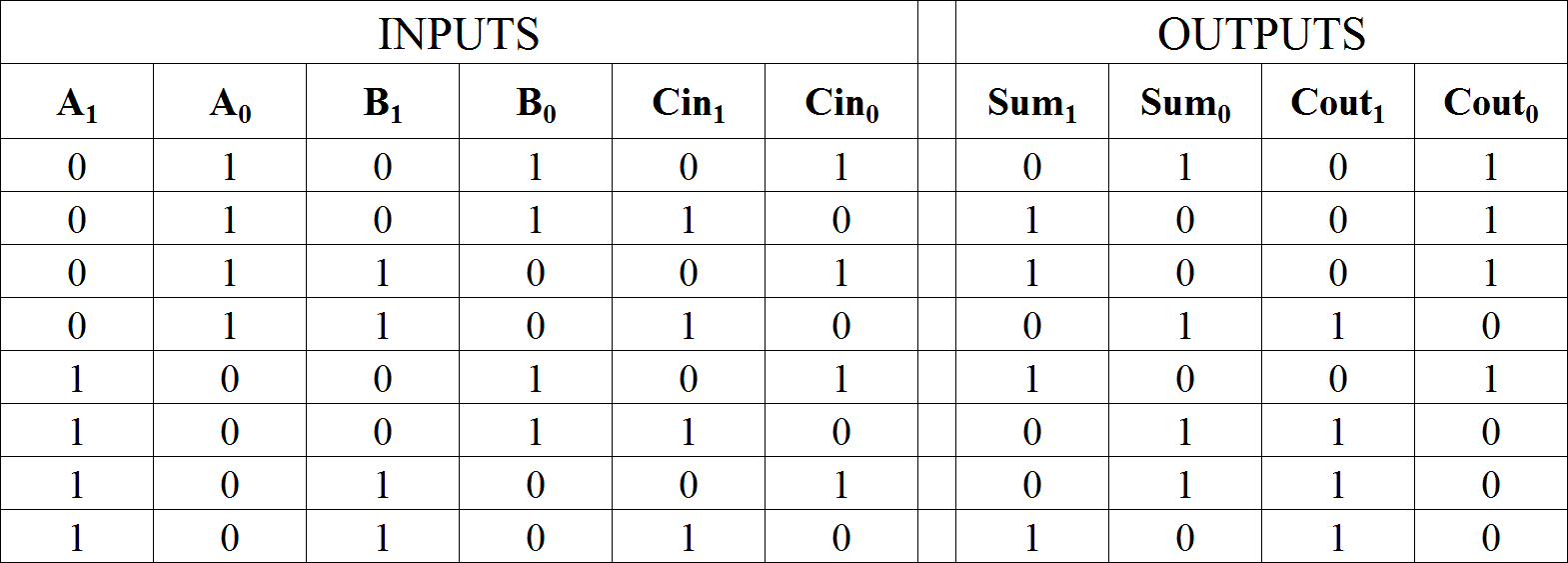}}
\caption{Logic table for the 1-bit full adder shown in Fig.~\ref{full adder}.}
\label{full adder table}
\end{figure}

While substantially more complex than the previous example, the NAND gate functions on the same principle of using locks and balances to implement logic. In this example, a set of inputs are connected to a set of locks. The inputs determine which side of each lock is locked. Another input (a clock signal in this example) is then used to actuate a main balance, the movement of which cascades through a series of additional balances and locks. Each balance moves either its top side or its bottom side, in accordance with the state of the locks to which it is connected. This results in a final output lock either having its top half or its bottom half move forward.

In Fig.~\ref{NAND-gate}, conceptually there are two binary inputs, A and B, and one binary output, X. The physical model used for the inputs and output is that a single 1-bit input or output is broken into two separate input or output links (we will refer to this as the ``two-input system'', although it is also called ``dual rail logic'' \cite{Buck_1956, Koller_1992, Teichmann_2011}). For example, the A input is composed of inputs $A_0$ and $A_1$. If the value of A is to be set to 0, the $A_0$ input moves. If the value of A is to be set to 1, the $A_1$ input moves. $A_0$ and $A_1$ are not permitted to move simultaneously (which is logically obvious since a value cannot be set to 0 and 1 at the same time). The same applies to the input B and output X.

Note that Fig.~\ref{NAND-gate} is just a single example, used as a proof due to the well-known universal nature of NAND. Using multiple NAND gates to create other logical functions may not always be efficient, and as previously mentioned, we have also shown that any of the standard 2-input logic gates can be implemented directly using locks and balances (see Appendix).

An example of a more complex logic function is shown in Fig.~\ref{full adder}. This device performs addition with carry on one bit (Fig.~\ref{full adder table}). These devices can be cascaded together to perform addition on more than one bit, as shown in Fig.~\ref{cascading adders}.

\section{Sequential Logic}
Having demonstrated that all necessary combinatorial logic can be created using only links and rotary joints (assembled into locks and balances), we turn to sequential logic. The outputs of the NAND gate depicted in Fig.~\ref{NAND-gate} are dependent solely upon the current inputs. Such a mechanism provides no way of storing previous inputs or the results of previous logical operations because the outputs return to zero when the inputs and clock return to zero; it provides no means for creating memory. To demonstrate the creation of a simple memory mechanism, we describe the design of a small shift register, again only using links and rotary joints.

A shift register can be built by combining locks and balances to create cells which are the logical equivalent of electronic flip-flops. Each cell of a shift register is related to its neighbor by virtue of relying upon a preceding or succeeding clock phase, as appropriate. This enables the copying and shifting of data through the shift register, rather than deterministically setting the contents of the entire shift register simultaneously. 

\begin{figure}
\centerline{\includegraphics[width=\columnwidth]{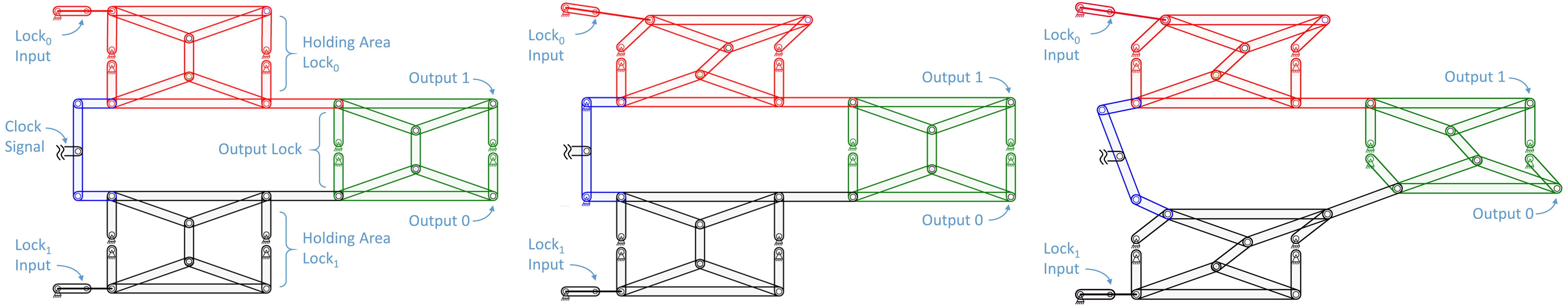}}
\caption{Left: Shift register cell in the (0,0) blank state; Center: Shift register cell with input (1,0) prior to clock actuation; Right: Shift register cell with input (1,0) after clock actuation}
\label{shift-register-cell-ABC}
\end{figure}

\begin{figure}
\centerline{\includegraphics[width=\columnwidth]{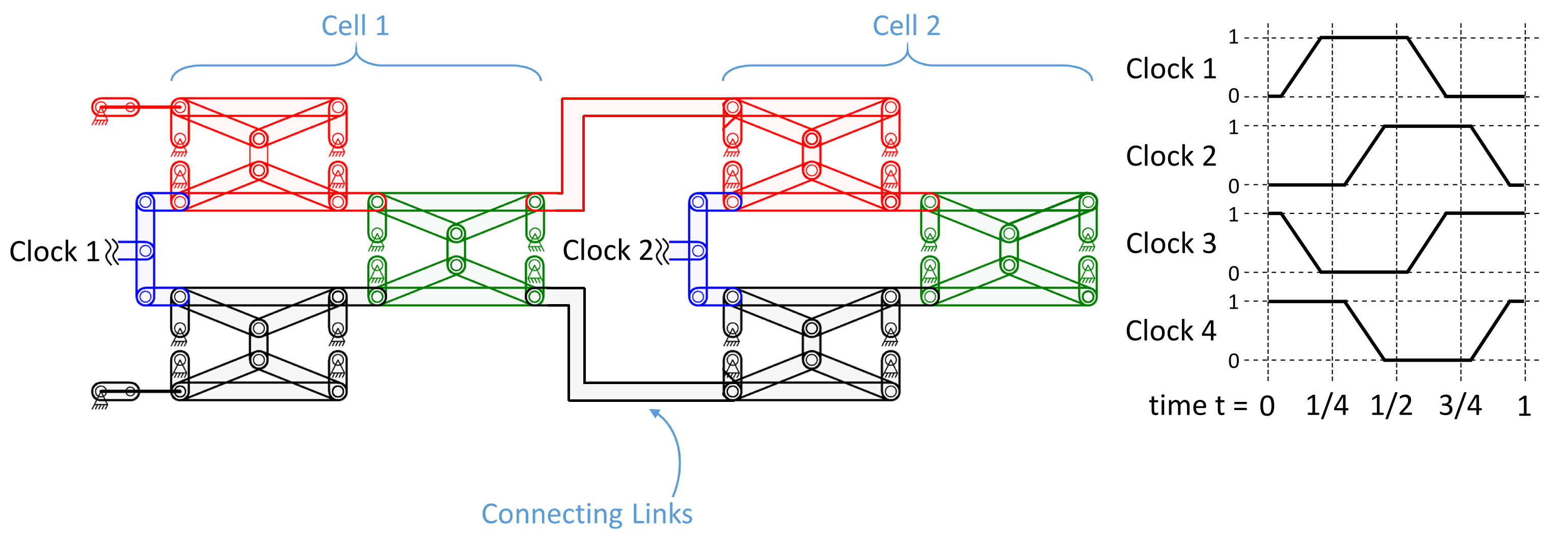}}
\caption{A two-cell shift register (left) shown with a plot of a four-phase clock cycle (right). In this example, the clock signal of Cell 1 is driven by Clock 1 and the clock signal of Cell 2 is driven by Clock 2.}
\label{two-cell-shift-register}
\end{figure}

\begin{figure}
\centerline{\includegraphics[width=5in]{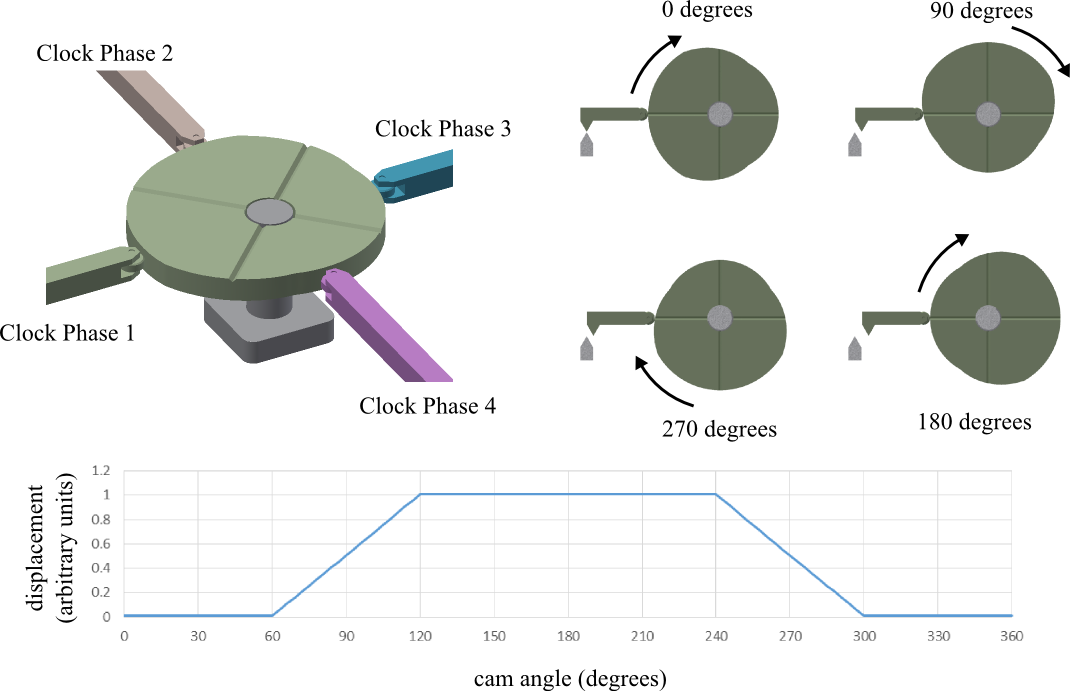}}
\caption{A suitable multiphase clock signal can be generated mechanically using cams and followers. Here, a four-phase clock signal is generated using four identical cams spaced 90 degrees out of phase. The four diagrams at the upper right show the cam at four rotations with only one of the four links. The collection of all four waveforms is shown in the right hand portion of Fig.~\ref{two-cell-shift-register}.}
\label{cam_w_timing}
\end{figure}

\subsection{A Shift Register Cell}
The left side of Fig.~\ref{shift-register-cell-ABC} depicts a single shift register cell with an input of (0,0), or what may be referred to as the blank state. Using the two-input system (i.e. dual rail logic), the other two possible inputs consist of zero, represented as (1,0), and one, represented as (0,1). The (1,1) state is generally not used, as that state would not permit either side of the balance to move. Note that the blank state suffers from the opposite problem: Both sides of the balance could move. However, the clock signal would not be driven to 1 while the locks are in the blank state, so this is not a practical problem.

Before delving into the actual use of shift register cells, a brief description of the major parts of a cell is in order. First, it is notable that the left-hand portion of a cell is identical to Fig.~\ref{lock-and-balance}, consisting of a balance which is connected to a top lock (Holding Area $Lock_0$) and a bottom lock (Holding Area $Lock_1$). The balance is actuated by a clock signal, again, just like Fig.~\ref{lock-and-balance}. The only difference is the addition of an extra lock, the Output Lock, which is connected to the outputs of the left side of the mechanism and in turn provides the final output for the cell. The easiest way to visualize how a cell works is to step through the movements.

Starting from the blank state, an input is set. Since this mechanism uses the two-input system, either the $Lock_0$ input is set to 1, or the $Lock_1$ input is set to one, but not both. If we assume that $Lock_0$ is set to 1, the movement results in the mechanism being in the state depicted in the center of Fig.~\ref{shift-register-cell-ABC}.

In the center of Fig.~\ref{shift-register-cell-ABC}, Holding Area $Lock_0$ has moved its upper side. This results in the lower side of that lock locking. However, since shift registers are dependent upon multiphase clocking, note that the input at $Lock_0$ has not yet had any effect on the Output Lock. For that to occur, the balance must be actuated, which does not occur until the next clock phase.

During the next clock phase, the balance is actuated (set from 0 to 1, in this case). Since $Lock_0$ has locked its lower half, which is connected to the upper side of the balance, upon actuation, the balance can only move its lower side. This movement propagates to the Output Lock, resulting in the state depicted on the right side of Fig.~\ref{shift-register-cell-ABC}.
 
The reason the left two locks are referred to as Holding Area Locks may now be apparent: They temporarily hold the input data prior to clock actuation. An input to these locks does not instantly result in an output at the Output Lock. Rather, the clock must actuate first, which results in copying the value, be it 0 or 1, from the holding area locks to the Output Lock.
Note that we adopt the convention that, with respect to the input locks, the top lock is associated with an input of 0, while on the Output Lock, the top half is associated with an output of 1. This is because, due to how the mechanism is diagrammed, when an input of 1 is set at the top input lock, an output of 1 ends up at the bottom half of the Output Lock, and vice-versa. The mechanism could be easily altered to change this, but as it is currently represented, from a naming perspective it is easiest to have the 0 input result in a 0 output, and the 1 input result in a 1 output.

Thus far, we have only needed two clock phases: On the first phase, the inputs are set, and on the second phase, the balance actuates and the inputs are copied from the holding locks to the Output Lock. However, in an actual system, additional phases would be used. The subsequent examples assume four-phase clocking.

\subsection{Connecting Cells}
\label{connecting}
Figure \ref{two-cell-shift-register} depicts a two cell shift register to illustrate how two cells would be connected and to explain how data would move from one cell to the next. In this figure, Cell 1 and Cell 2 are each equivalent to the mechanism depicted in Fig.~\ref{shift-register-cell-ABC}. The connecting links connect the output from Cell 1 to the inputs to Cell 2. Both cells also have a connection to a clock signal created by a mechanism such as that shown in Fig.~\ref{cam_w_timing}.

The operation of a single cell has already been described. Now, demonstrating how Cell 1 passes data to Cell 2 will illustrate the function of a minimal shift register. The sequence of events is as follows:

\begin{enumerate}
\item At time $t = 0$, the clock input for Cell 1 has been set to 0, and the data inputs have been set for Cell 1.
\begin{itemize}
\item Either the upper or lower lock of Cell 1 is locked, depending on which input was set to 1.
\end{itemize}
\item During the transition from $t = 0$ to $t = 1/4$, the clock signal for Cell 1 is set to 1.
\begin{itemize}
\item This results in the unlocked side of the Cell 1 balance moving, which in turn moves the upper (if the input value was 1) or the lower (if the input was 0) half of the output lock.
\item Cell 1's output lock in turn moves the appropriate connecting link, locking one of Cell 2's holding area locks. This has effectively copied the data from Cell 1's output lock into one of Cell 2's holding area locks.
\item Note that the output lock of Cell 2 still has not moved.
\end{itemize}
\item During the transition from $t = 1/4$ to $t = 1/2$, the clock signal for Cell 2 is set to 1.
\begin{itemize}
\item This copies the data from Cell 2's holding area locks into Cell 2's output lock.
\end{itemize}
\item During the transition from $t = 1/2$ to $t = 3/4$, the clock for Cell 1 is reset to 0.
\begin{itemize}
\item As a result, Cell 1's output lock, and hence the connecting links to Cell 2, retract to the 0 position (regardless of the values still stored in Cell 1's holding area cells). 
\end{itemize}
\item During the transition from $t = 3/4$ to $t = 1$, the clock for Cell 2 is reset to 0.
\end{enumerate}

\begin{figure}
\centerline{\includegraphics[width=\columnwidth]{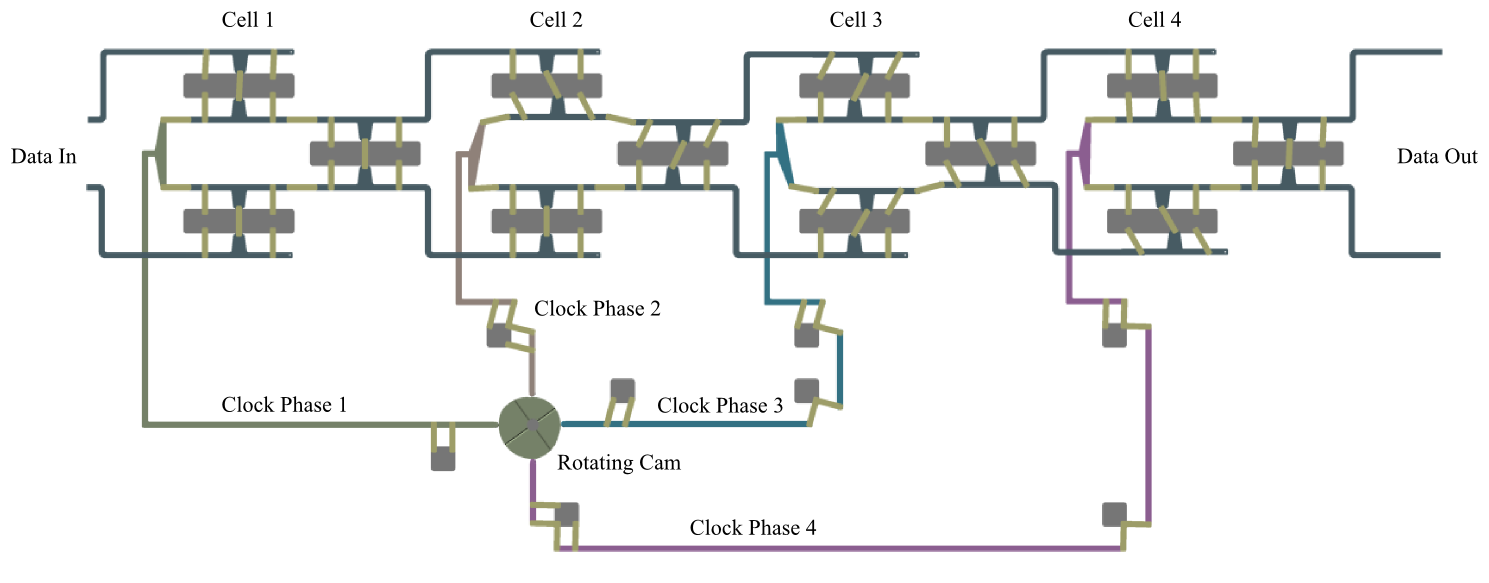}}
\caption{A 4-cell shift register driven by a four-phase clock, shown at time $t = 3/4$. The last three cells are set to state 1 and the first cell in the blank state. Animations of this mechanism operating in forward and reverse are available at \cite{merkle_online_2018_a} and \cite{merkle_online_2018_b} respectively. Solid models available online at \cite{github_models}.}
\label{four-cell-shift-register}
\end{figure}

This cycle then repeats itself as new data is input into Cell 1. In step 2 above, it is noted that the output lock of Cell 2 still has not moved. This behavior allows shift register cells to store previous data. This is a key difference between combinatorial and sequential logic. The state of the NAND gate described previously is completely determined by the current data and clock inputs. However, this is not true of shift registers. For example, a shift register with four cells that is driven by a four-phase clock is able to store one bit. One of the four clock phases will always be active for one of the four cells, and thus one of the four cells will contain information. When combined with transition logic to handle input and write enable, this allows the mechanism to act like a mechanical analog of an electronic flip-flop, thereby forming the basis for memory storage. To aid in the visualization of how this mechanism works, a 4-cell shift register is diagrammed in Fig.~\ref{four-cell-shift-register}, and an animated version is available online \cite{merkle_online_2018_a}  (see also Fig.~\ref{state machine}).

\begin{figure}
\centerline{\includegraphics[width=3in]{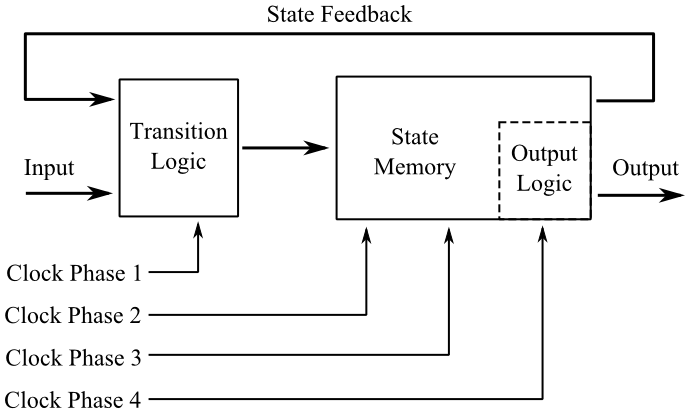}}
\caption{Block diagram of a generic Moore machine adapted for a four-phase clock. The state memory is implemented as a chain of shift register cells, similar to those shown in Fig.~\ref{four-cell-shift-register} and Fig.~\ref{cascading adders}.}
\label{Moore machine}
\end{figure}

\begin{figure}
\centerline{\includegraphics[width=\columnwidth]{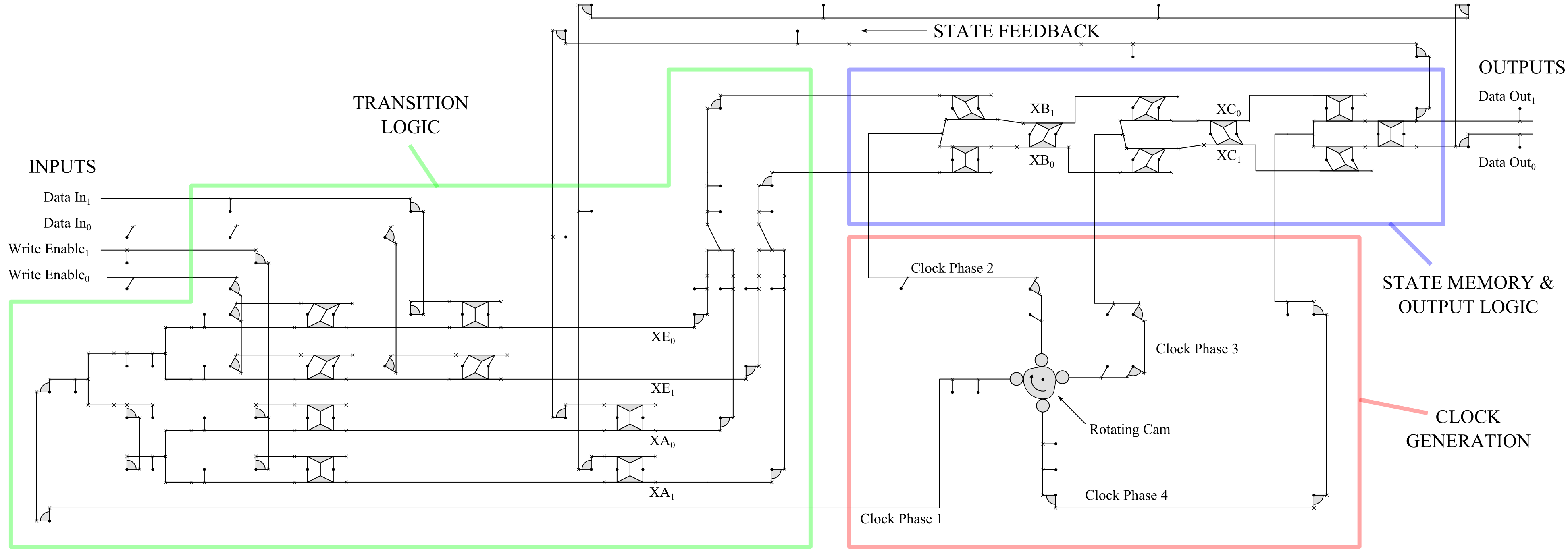}}
\caption{A simple state machine. The main components are highlighted for comparison with Fig.~\ref{Moore machine}. This state machine implements the transition table shown in Fig.~\ref{state transition}. A table of the schematic symbols used in this drawing is provided in the Appendix (Fig.~\ref{schematic symbols}).}
\label{state machine}
\end{figure}

\begin{figure}
\centerline{\includegraphics[width=5in]{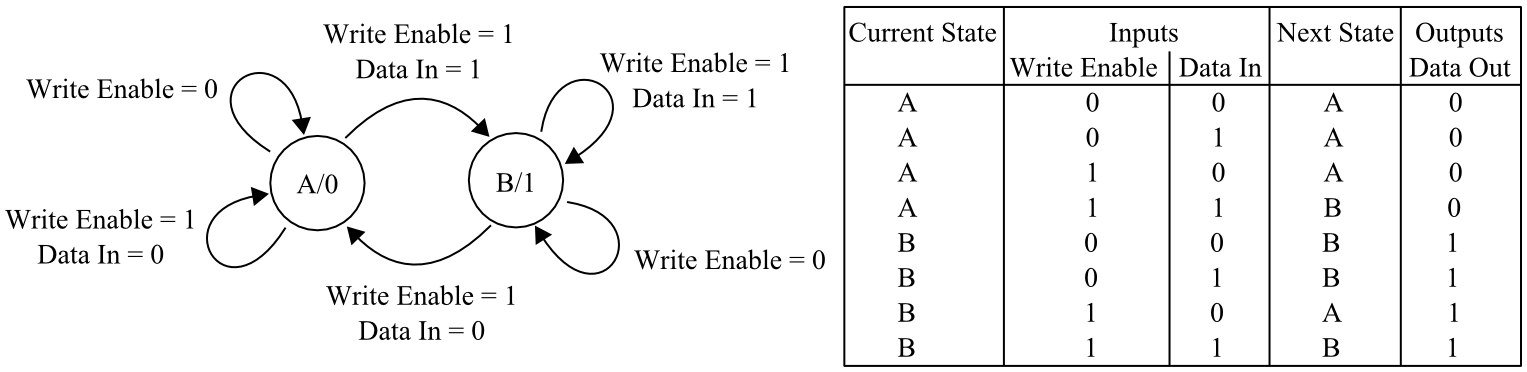}}
\caption{State transition diagram and table for the simple state machine of Fig.~\ref{state machine}.}
\label{state transition}
\end{figure}

\begin{figure}
\centerline{\includegraphics[width=\columnwidth]{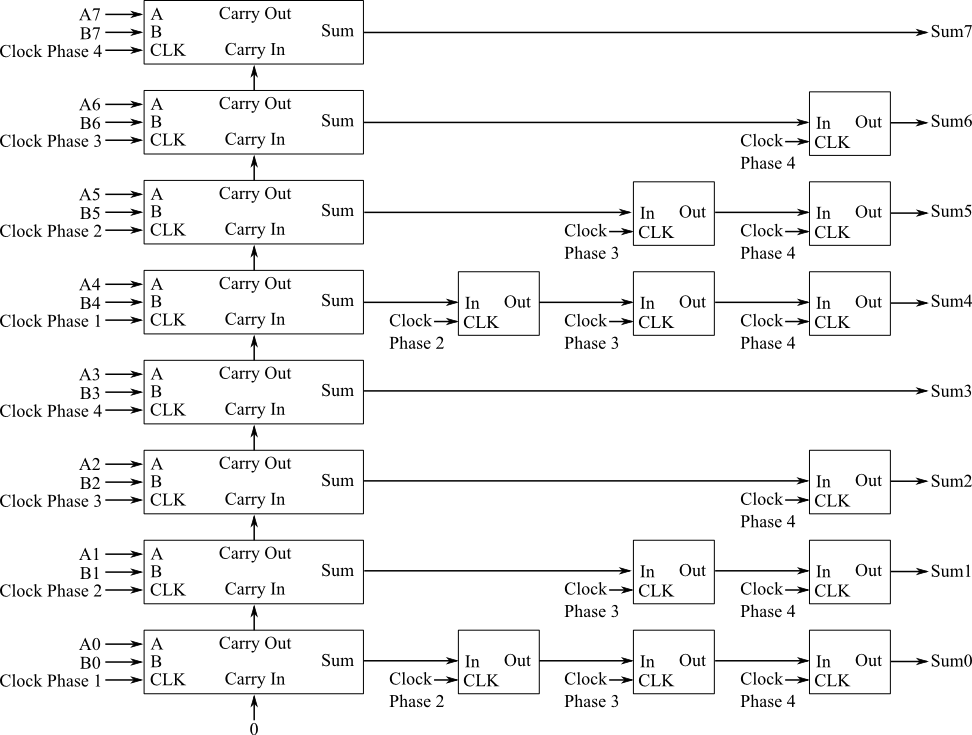}}
\caption{Eight 1-bit full adders (wide blocks) are cascaded using ripple carry. As described in Section \ref{connecting}, multiple blocks can be cascaded using a four-phase clock. Narrow blocks are shift register cells, which form a delay line that stores portions of the results during computation. The final result appears on the outputs (right side) after two full clock cycles. }
\label{cascading adders}
\end{figure}

\begin{figure}
\centerline{\includegraphics[width=4.5in]{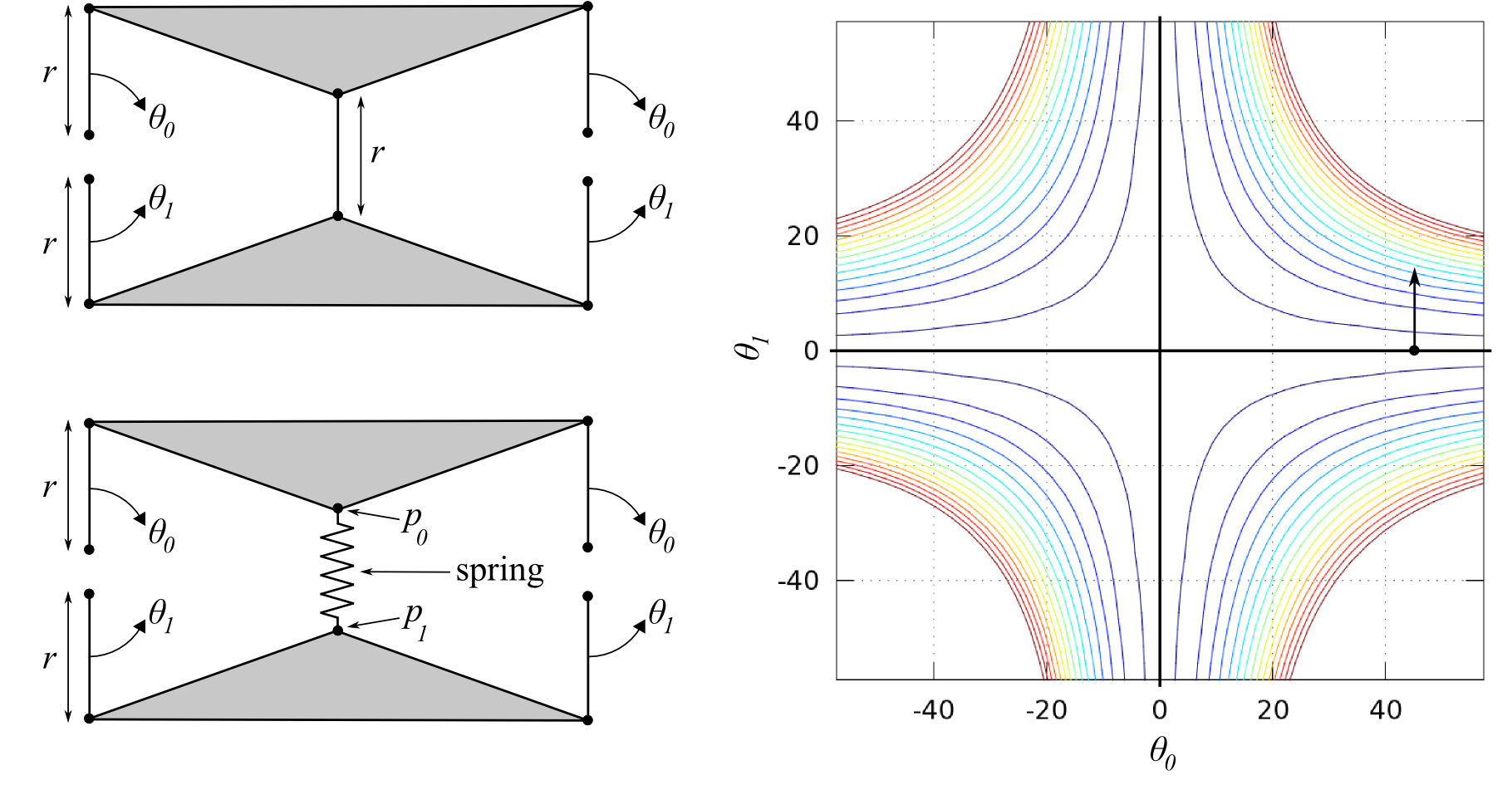}}
\caption{Analyzing lock holding force. In the ideal case (upper left) the lock has two overlapping sets of solutions: $\theta_0 = 0, \theta_1 \ne 0$ and $\theta_0 \ne 0, \theta_1 = 0$, with a kinematic branch point \cite{Gosselin_1990,Park_1999} at $\theta_0=0, \theta_1=0$. Link flexibility can be modeled by replacing one of the links with a spring (lower left). The contour plot shows level sets of spring energy as a function of input angles. In the ideal case, an arbitrarily large force can be locked without effecting the other input. In the more realistic case including link flexibility, a small holding force on one input can hold a large force on the other input.}
\label{locking force}
\end{figure}

\subsection{Power Source and Clock Generation}
The clock signal essentially functions as a synchronized power source. Multiphase power/clock signals were proposed for electronic low-power adiabatic logic systems in \cite{Athas_1994,Moon_1996}. An example multiphase clock signal is shown in Fig.~\ref{two-cell-shift-register}. The amplitude and shape of the clock waveform are somewhat arbitrary, provided that 
\begin{enumerate}
\item{There is sufficient overlap between adjacent clock phases when they are active: there must be a time when two adjacent clock phases are both active.}
\item{There is sufficient dwell time: there must be a duration of time when a clock phase is completely inactivated.}
\end{enumerate}
There are many ways that a suitable multiphase clock signal can be generated, both purely mechanically and otherwise. Examples include but are not limited to cams and followers (see Fig.~\ref{cam_w_timing}, Fig.~\ref{four-cell-shift-register}, and Fig.~\ref{state machine}); linkage-based dwell mechanisms; devices based on crankshafts, springs, and mechanical stops; and electrically driven MEMS comb actuators.

\subsection{Finite State Machines}
Combinatorial and sequential logic can be combined to build finite state machines. A block diagram of a generic Moore machine adapted for a four-phase clock is shown in Fig.~\ref{Moore machine}. Transition logic is implemented using the method described in Section \ref{UCL}. State memory is implemented using a chain of shift register cells as described in Section \ref{connecting}. A detailed example of a state machine is shown in Fig.~\ref{state machine}. This machine functions essentially as a one bit memory, as described in the state transition table in Fig.~\ref{state transition}.

\subsection {Scaling to Larger Systems}
\label{scaling}

The method for dealing with scaling is to isolate individual cells in a shift register (or more generally, individual clocked logic elements, such as Fredkin gates) by using the four-phase clock. That is, forces along signal lines cannot move beyond two cells before encountering a locked lock, preventing any accumulation of forces over long distances. At any point in time, a cell is either blocked from transmitting forces to adjacent cells, or can transmit forces only to one other cell (either the predecessor or the successor).  An example of a scaled-up system that avoids force accumulation is shown in Fig.~\ref{cascading adders}. 

Un-clocked mechanical logic systems allow forces to accumulate over multiple logic levels, which can be a problem. It is sometimes advantageous to use logic systems that span as many logic levels as possible without an intervening clock to isolate the logic levels. Babbage, in his original design for an adder, used this idea to propagate the carry along the full length of the register with one mechanical motion \cite{Bromley_1983}, in a manner analogous to what is today called a ``carry lookahead" adder. This is in contrast to the mechanism in Pascal's calculator, which used stored energy to propagate a carry along an indefinite length of wheels \cite{Touretzky_2015}, analogous to what is now called ``ripple carry". Babbage's design is susceptible to force accumulation, whereas Pascal's is not. 

Force accumulation in link logic systems can be mitigated because each lock acts as a mechanical amplifier, with a small force controlling a much larger one (see Fig.~\ref{locking force}). In the ideal case of rigid links and perfect pivots, the lock is kinematically constrained such that an arbitrarily large force on the locked input will not be transmitted to the other (moved but unlocked) input. This behavior is enabled by a kinematic branch point \cite{Gosselin_1990,Park_1999} at $\theta_0=0, \theta_1=0$ that allows for the discrete switching between the two sets of kinematic solutions. 

Actual mechanisms will of course have some flexibility in the links. This can be modeled in a simple way by replacing one of the links with a spring, of stiffness $k$ and equilibrium length $r$, as shown in the lower left of Fig.~\ref{locking force}. The contour plot on the right of Fig.~\ref{locking force} shows the level sets of spring energy $V(\theta_0, \theta_1) = (k/2) ( ||\vec{p_0} - \vec{p_1}|| - r)^2$. A test point in the jointspace is shown as a dot near $\theta_0=45$ degrees, $\theta_1 = 0$ degrees. Visual inspection of the contours near this point show that a small motion in $\theta_1$ will result in a large restoring force in the direction of $\theta_0$ and a very small deflecting force in the direction of $\theta_0$. This confirms that, even in a physical lock with flexible links, a small holding force in one input can hold a large force on the other input, provided the activated input has moved far enough.

\subsection{High-density Memory}
\begin{figure}
\centerline{\includegraphics[width=\columnwidth]{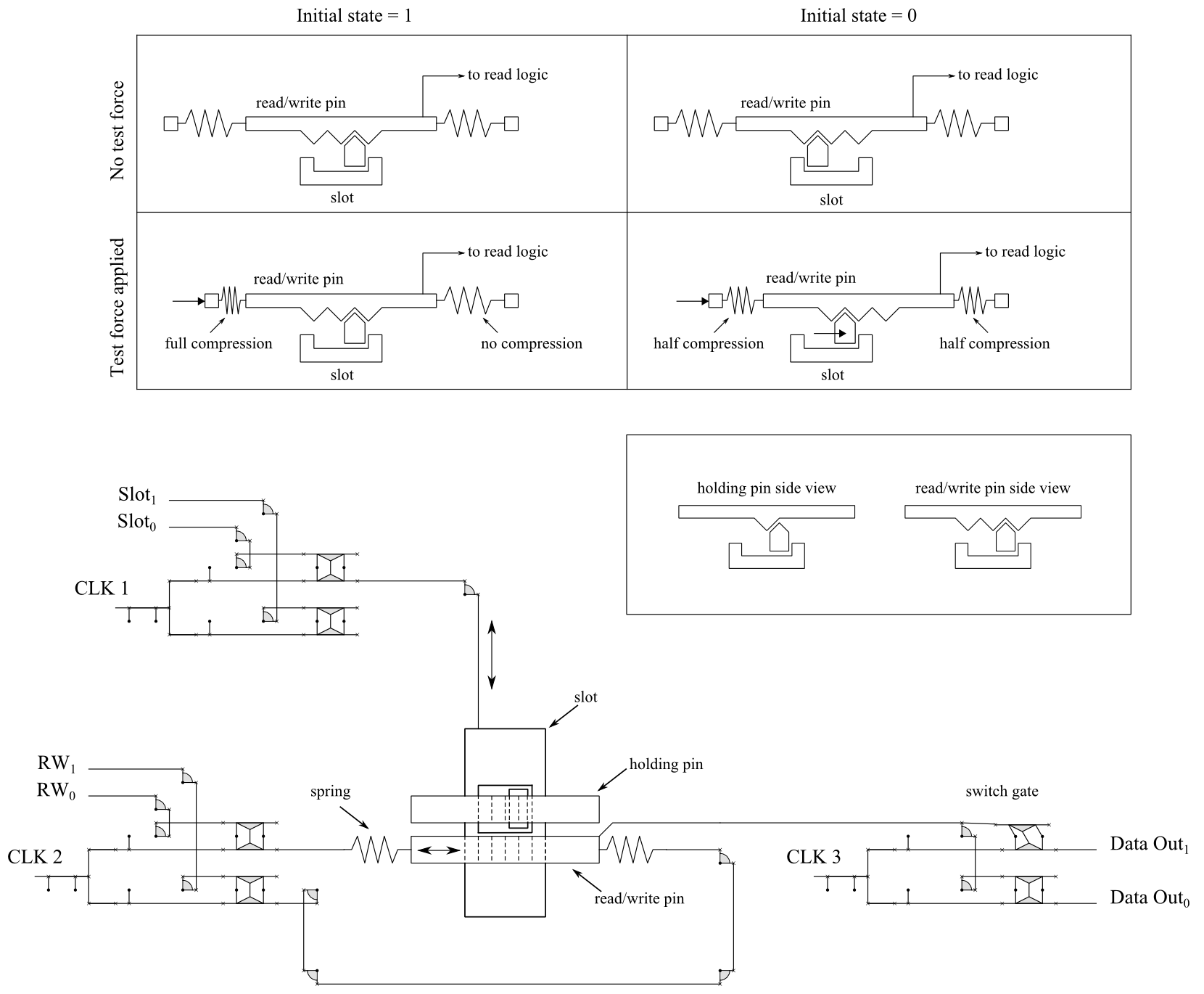}}
\caption{Logic circuitry for read and write operations on a high density memory storage cell.}
\label{SHDMC}
\end{figure}

The state machine in Fig.~\ref{state machine} is closely equivalent to a D flip-flop -- it stores one bit of information on every clock cycle. This circuit is suitable for use in state machines, but it is not an efficient use of gates for high-density memory applications. While it departs from the links-and-pivots-only schema, the mechanism shown in Fig.~\ref{SHDMC} uses far less gates to store a bit of information. The tradeoff is that the read/write process is somewhat slower and more involved.

\subsection{Reversibility}
\begin{figure}
\centerline{\includegraphics[width=\columnwidth]{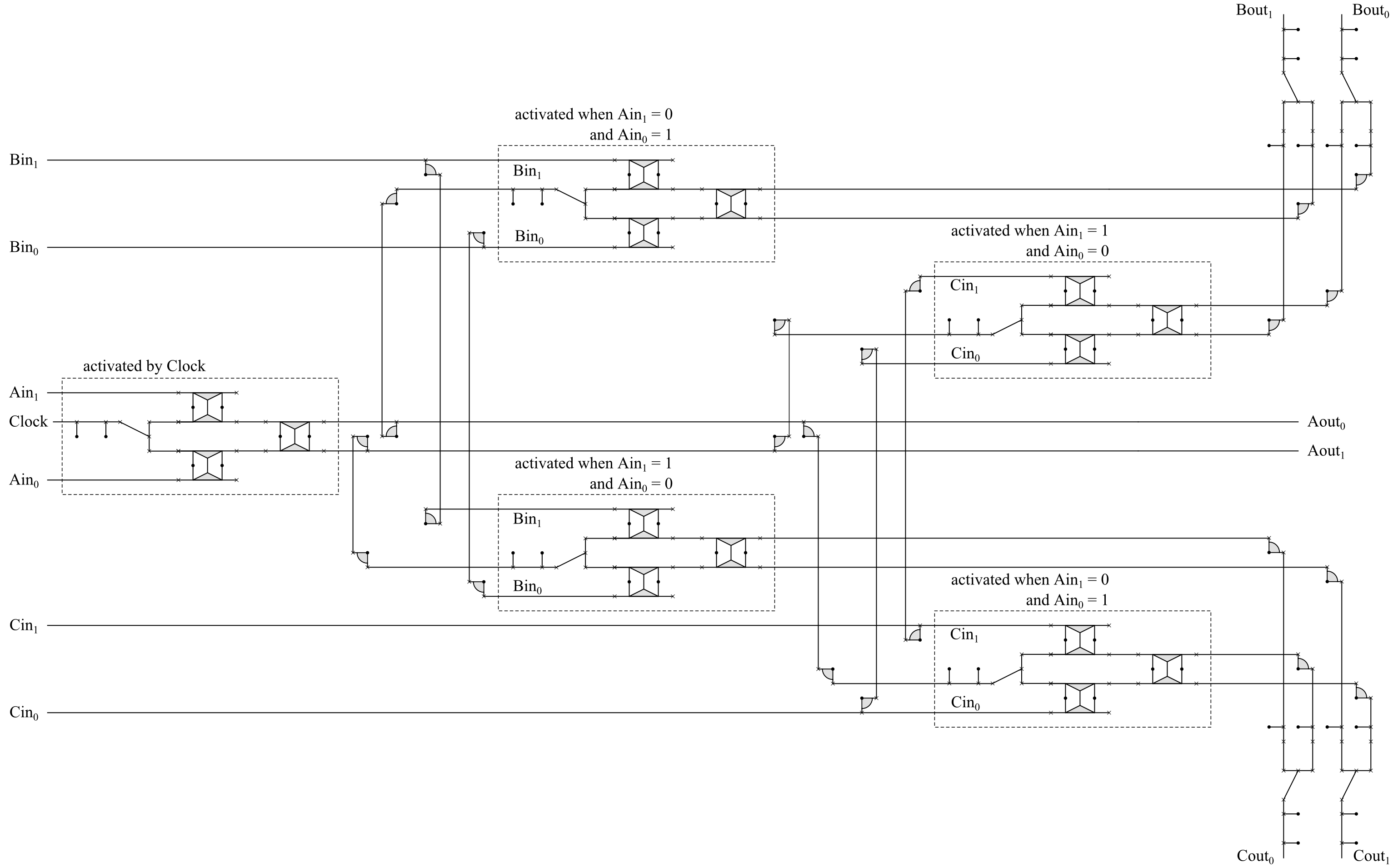}}
\caption{A balance- and lock-based Fredkin (CSWAP) Gate, using a two-link per bit design (i.e. dual rail logic). This mechanical logic gate is logically and physically reversible. The logic table implemented is shown in Fig.~\ref{fredkin table}. }
\label{fredkin_v01}
\end{figure}

\begin{figure}
\centerline{\includegraphics[width=4in]{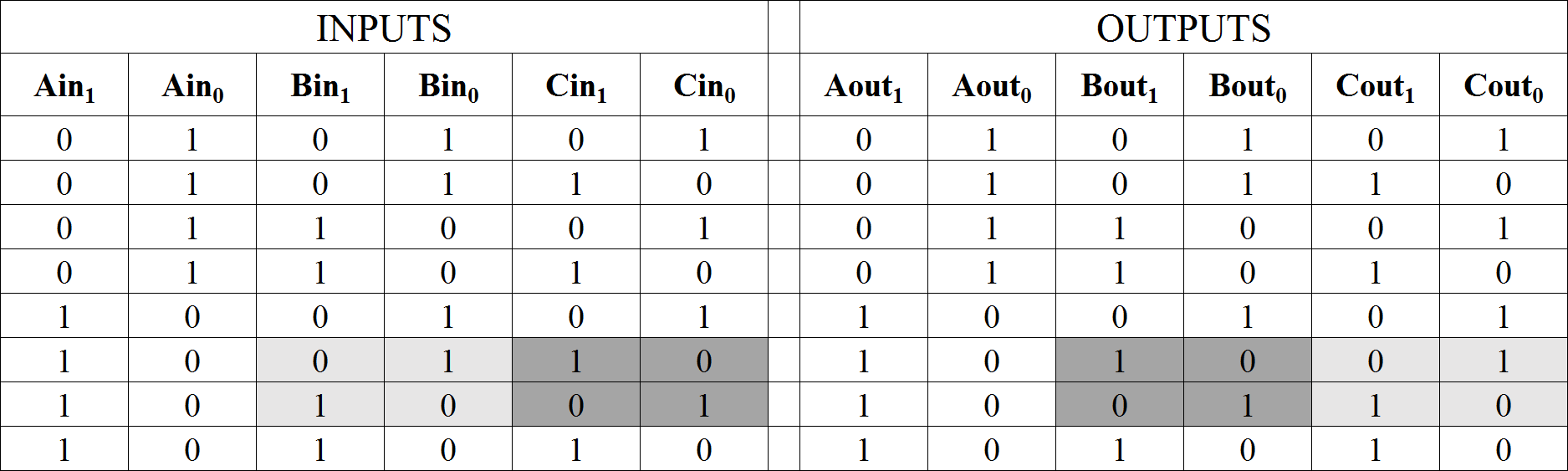}}
\caption{Logic table for reversible Fredkin (CSWAP) gate (see Fig.~\ref{fredkin_v01}). Identical portions of the inputs and outputs are highlighted in the same color as a guide for the eye. }
\label{fredkin table}
\end{figure}

Close inspection of the shift register chain in Fig.~\ref{four-cell-shift-register} reveals that, if the clock generating mechanism is operated in reverse, information will propagate backward through the chain, from outputs to inputs. This is a characteristic and intentional feature of the mechanical link logic architecture. It is well known that reversible computers can be constructed from reversible logic gates such as the Fredkin gate \cite{bennett82,Wenzler_2014}. Figure \ref{fredkin_v01} shows how physically reversible Fredkin gates can be constructed from links and rotary joints (logic table in Fig.~\ref{fredkin table}). Reversibility can be exploited to create computers with extremely low energy dissipation. This potential application is discussed further in Section \ref{APM} .

\section{Implementations and Applications}
Mechanical computers have potential applications at the meso-, micro-, and nano-scales. Applications at the meso- to micro-scale include such disparate areas as soft robotics \cite{raney2016stable,Ion_2017} and failsafe devices resistant to radiation and electromagnetic pulse (EMP) \cite{plummer1998history}. Mechanical computers constructed from nano-scale components could potentially dissipate much less energy than conventional CMOS devices, while providing comparable computing performance. Manufacturing at such small scales is challenging. The architecture described in this paper, consisting mainly of links and rotary joints, could be well-suited for such sizes since it does not require fabricating a large set of complicated components. In particular, a molecular version of this architecture could use stiff covalently-bonded nanotubes for the links and single bonds for the joints.

\subsection{Input and Output}
\begin{figure}
\centerline{\includegraphics[width=\columnwidth]{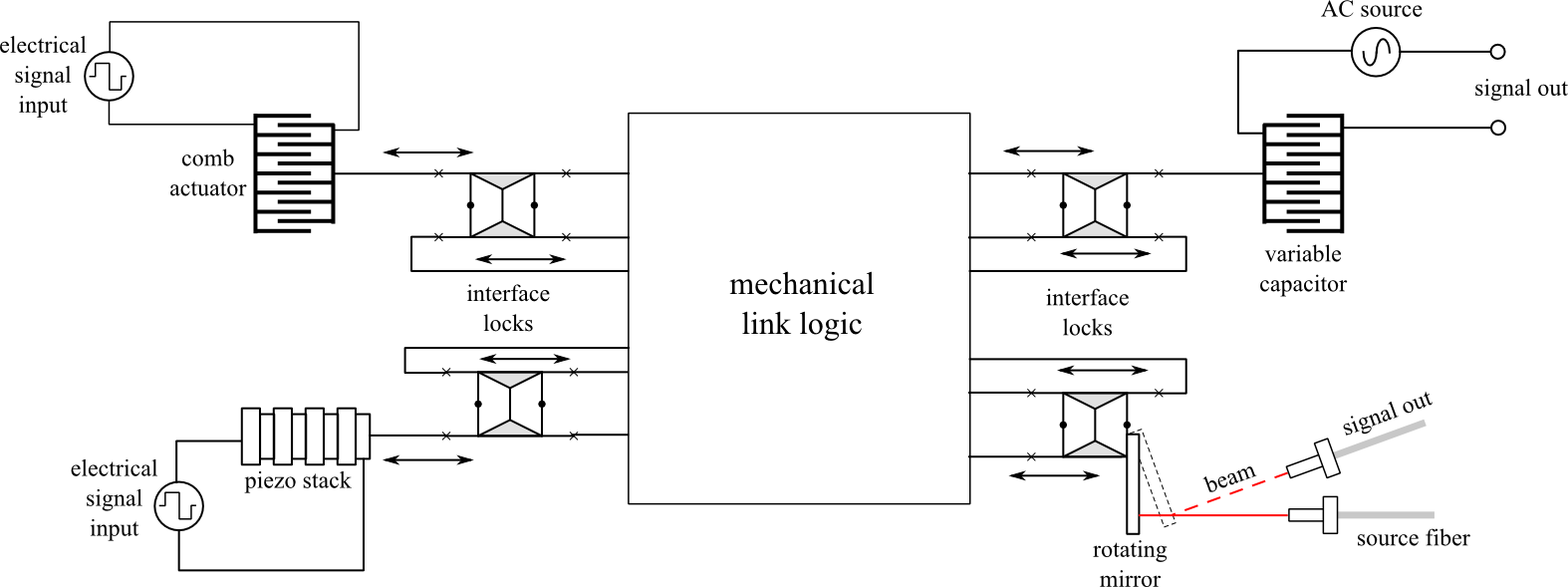}}
\caption{On the left, two concepts for inputting signals to the mechanical computer are shown; two concepts for outputting signals are shown on the right. }
\label{input_output_concepts}
\end{figure}

In many potential applications where a mechanical computer is constructed for its particular advantages (low energy dissipation, radiation hardness, etc) it may still be necessary to interface the mechanical elements to conventional electronic devices. Figure \ref{input_output_concepts} illustrates a number of concepts for transferring information in and out of a mechanical computing device. In the top left, a conventional electrostatic MEMS comb actuator is used to move the (presumably very small) mechanical input links. In the lower left the input links are moved by a piezoelectric actuator. On the top right, mechanical links are used to move a MEMS variable capacitor, the value of which is read off by conventional electronics. On the lower left an optomechanical scheme is shown where the output links move a mirror to modulate an optical signal interfaced to standard optoelectronic components. 

\begin{figure}
\centerline{\includegraphics[width=\columnwidth]{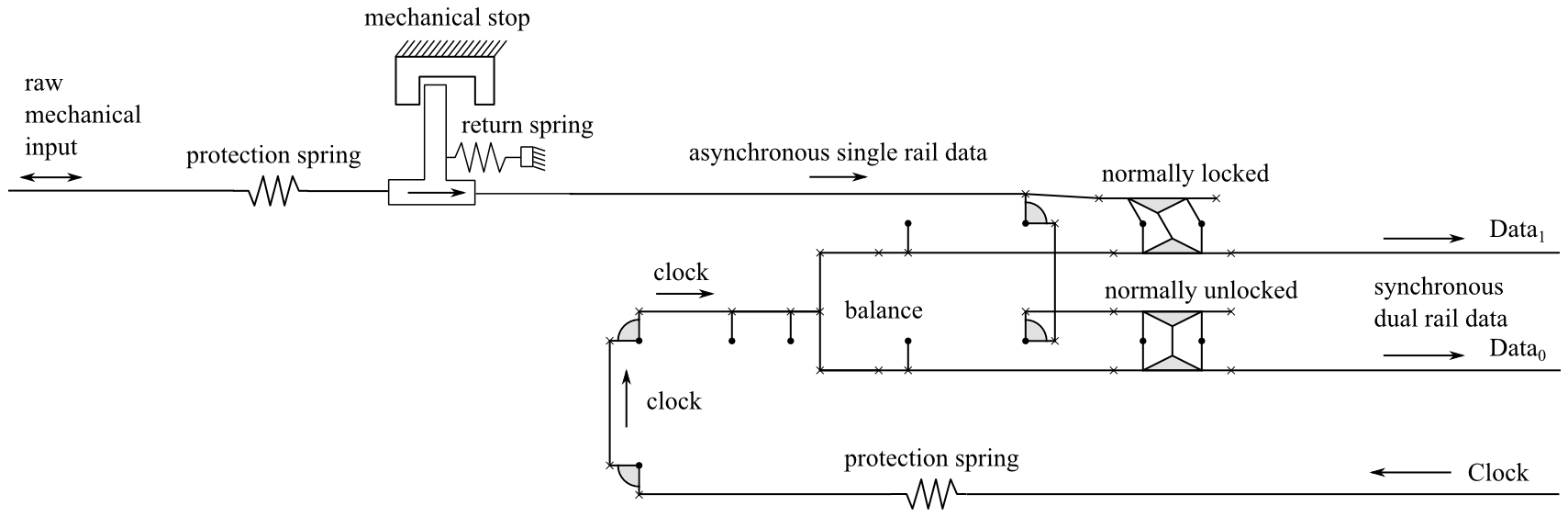}}
\caption{Asynchronous, noisy signals (such as from a mechanical sensor) can be conditioned so that synchronous dual rail mechanical logic gates can process them. }
\label{signal_conditioning}
\end{figure}

Typical input signals will be noisy and asynchronous, with amplitudes exceeding the allowed range of motion for the logic gates. These signals can be conditioned using mechanisms such as that shown in Fig.~\ref{signal_conditioning}.

\subsection{Macroscopic Components}
\begin{figure}
\centerline{\includegraphics[width=3in]{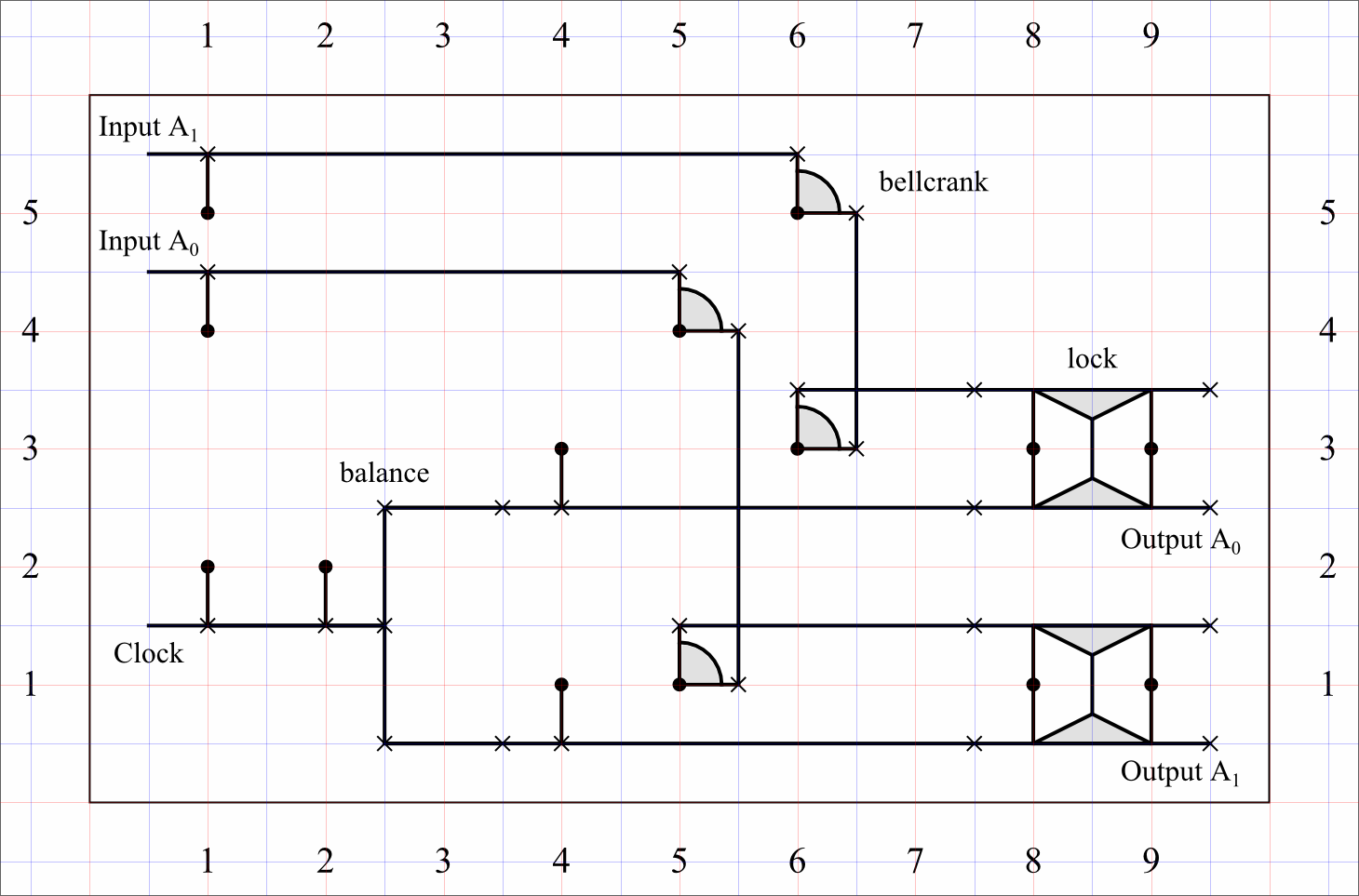}}
\caption{Schematic layout for a simple test mechanism containing a balance, two locks, and signal routing mechanisms.}
\label{simple testbed}
\end{figure}

\begin{figure}
\centerline{\includegraphics[width=5.0in]{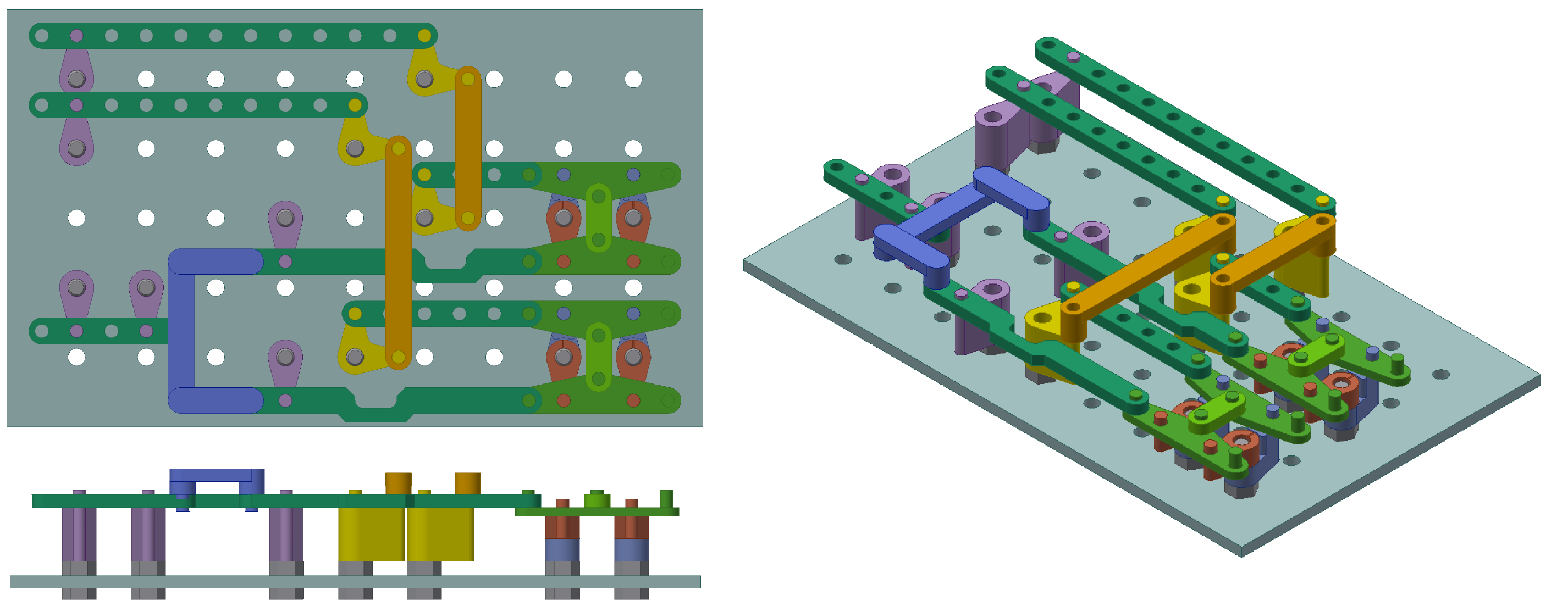}}
\caption{CAD model of the test system shown in Fig.~\ref{simple testbed} implemented with 3D printed components assembled over standard hardware and pegboard. Solid models available online at \cite{github_models}.}
\label{macro implementation}
\end{figure}

A straightforward option for creating macroscale mechanical link logic devices is shown in Fig.~\ref{simple testbed} and Fig.~\ref{macro implementation}. Standard pegboard material is used as the foundation, with 0.25 inch on 1.0 inch centers. Off-the-shelf bolts serve as the pivots, and the remaining links and locks can be made by 3D printing. While this is an effective method for quick demonstrations and educational purposes, it is unlikely to find much practical use outside these niche applications. 

\subsection{Flexure-based Designs}
\begin{figure}
\centerline{\includegraphics[width=5.5in]{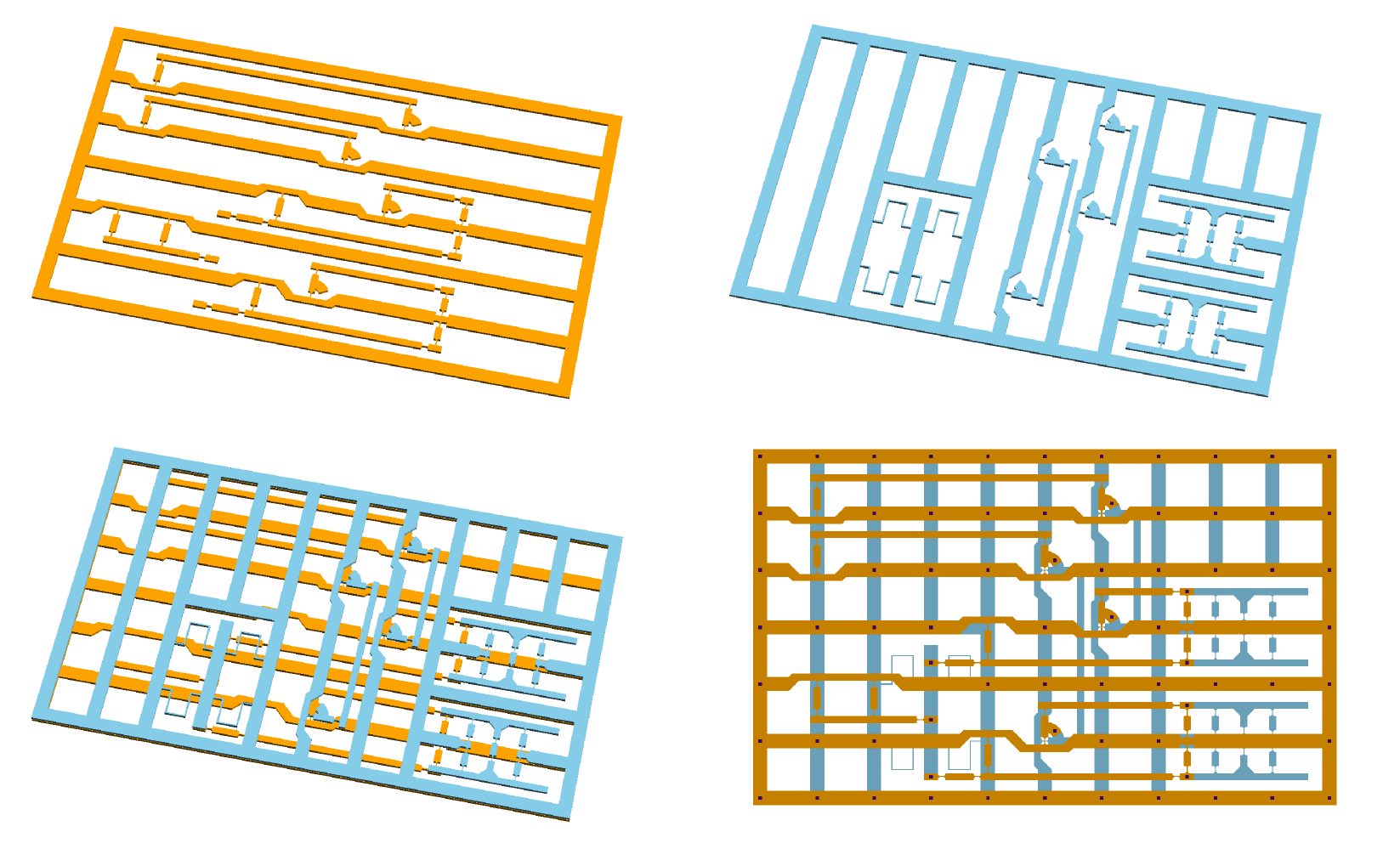}}
\caption{A flexure-based implementation of the system shown in Fig.~\ref{simple testbed}, made of two or three stacked layers (a second outer layer can be added to make a three-layer assembly for increased rigidity) . Interlayer bonds are shown as dark dots in the lower right image. Solid models available online at \cite{github_models}.}
\label{flexure implementation}
\end{figure}

Flexure joints provide an alternative implementation with similar general performance to pivots. Flexures have the advantage that in many cases, particularly with MEMS, they are easier to fabricate and often more reliable than fully functional pivot joints. A conceptual design of a flexure-based mechanical link logic system is shown in Fig.~\ref{flexure implementation}. A systematic method of design allows all necessary locks, balances, bell cranks, support links, and transmission links to be implemented in only two layers of material. These layers are shown in Fig.~\ref{flexure implementation}, with solid models avaiable online at \cite{github_models}. Spot-welds or rigid bonding between layers, as often used in flip-chip MEMS, holds the layers together at grid points. Each layer is a monolithic pure 2D pattern, making this approach well suited for conventional microfabrication techniques such as LIGA, silicon micromachining, or high-resolution additive manufacturing. 

As an example of what could be possible with conventional MEMS technology, consider that the minimum feature size of the popular Multi-User MEMS Processes (MUMPs) commercial program is two microns. If the flexures are two microns in width, a MEMS implementation of Fig.~\ref{flexure implementation} would cover an area of $640 \times 1070$ microns. Counting Fig.~\ref{flexure implementation} as the equivalent of two transistors, a MEMS system on a silicon die 2.8 cm square could contain the mechanical equivalent to 2,200 transistors, which is the transistor count in the 4-bit Intel 4004 CPU.

\subsection{Atomically Precise Manufacturing}
\label{APM}
Recently there has been some resurgence of interest \cite{Forrest_2015,service16,kassem2017} in molecular machines created by atomically precise manufacturing \cite{drexler92}. Apart from their ultimate miniaturization, molecular machines would offer the advantages of very low friction and zero wear. Mechanical computers constructed from molecular-scale atomically-precise components would be highly desirable because of their potential for combined high performance and low energy dissipation.

\subsubsection{Drag on Molecular Rotary Joints}
A key performance metric of computers is their energy dissipation. One contribution to dissipation is friction at the rotary joints in each logic gate. Due to the joint's small frictional drag, mechanical computers constructed from them can, in principle, dissipate orders of magnitude less power than conventional semiconductor computers, while still operating at relatively high speeds.

\begin{figure}
\centering  
\includegraphics[width=5.5in]{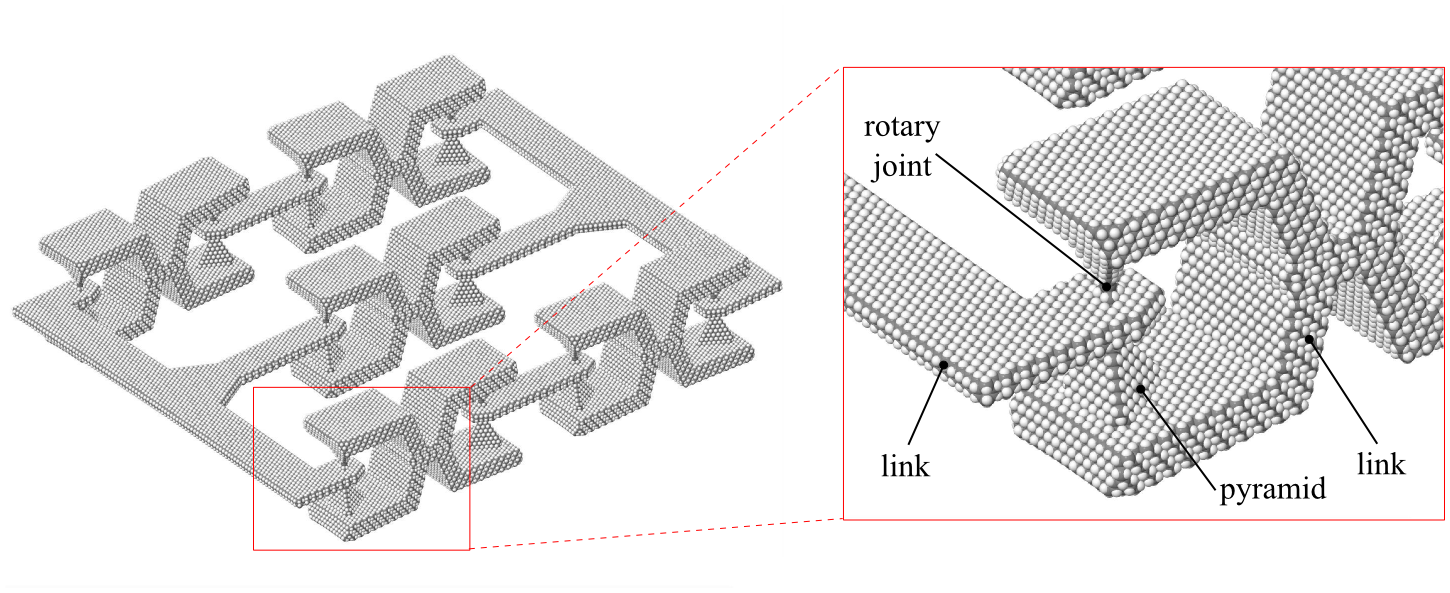}
\caption{Part of a molecular mechanical logic gate. This molecular machine consists of 120695 atoms, 87595 carbon and 33100 hydrogen, and occupies a volume of about  $27\,\nanometer \times 32\, \nanometer \times 7\, \nanometer$. The nine rigid links are connected to each other via a pair of rotary joints. Atomic structure file available online at \cite{github_models}.}
\label{logic gate}
\end{figure}

For instance, Fig.~\ref{logic gate} shows a lock containing bonded rotary joints \cite{merkle16}. Operating this lock involves rotation at the joints by up to $\Delta \theta \approx 1\,\radian$. The model system analyzed in \cite{Hogg_2017} is an excerpt of the links and joints shown in the closeup on the right of Fig.~\ref{logic gate}. From \cite[Eq. 2]{Hogg_2017}, this rotation dissipates about $2.4\times 10^{-27}\,\joule$ per rotary joint when operating at $f=100\,\megahertz$. Operation of the complete lock, involving multiple joints, would dissipate about an order of magnitude more.
This dissipation is several orders of magnitude smaller than  $\BoltzmannConstant T = 4.1 \times 10^{-21} \joule$ at $T = 300\, \Kelvin$. Thus at this operating speed, the rotor frictional dissipation per logic operation would be far below $\BoltzmannConstant T$. 

Fully exploiting the rotor's low dissipation for a computer requires avoiding other sources of dissipation. For instance, the clock system driving the logic gates could store and release energy to the extent it might be needed to move up or over any potential barriers that might be encountered~\cite{Hogg_2017}. More fundamentally, the computer would need to use reversible logic gates based on this rotary joint, to avoid the minimum $\BoltzmannConstant T \log 2$ dissipation from each logically irreversible bit operation predicted theoretically~\cite{frank05,bennett82,landauer61} and observed experimentally~\cite{berut12,hong16}. Such a machine could perform arbitrarily many computing operations per dissipated joule by operating at a sufficiently slow clock speed. The lowest possible energy dissipation would be aided by reducing or even eliminating potential barriers in those mechanical degrees of freedom involved in normal device operation.

A common performance metric for semiconductor devices is (energy dissipated per operation) multiplied by (time per operation). For dissipation due to velocity-dependent friction, as is the case for rotary joints, dissipation is proportional to speed, hence 1/time for operation requiring a fixed amount of motion, e.g., 60-degree rotation for the gate operations described in this paper. Hence the energy*time performance measure is a constant. From \cite[Eq. 2]{Hogg_2017}, this constant is $\Edissipated t = \Kdrag \phi^2$ for rotation by angle $\phi$. Supposing a logic operation corresponds to rotating about 10 joints by about a radian, this energy-time product is about $10^{-34}\,\joule \, \second$, using $\Kdrag$ from \cite{Hogg_2017}.

For a machine using multiple rotary joints, such as a computer, interactions between nearby rotors could affect the dissipation. This could constrain how closely rotors can be placed before interactions significantly increase dissipation. 
An experimental study of arrays of molecular rotors on a surface~\cite{zhang16} illustrates the effect of such interactions.

\subsubsection{Inertial Effects}
The inertial effects of the moving lock components would be small. Consider the lock shown in Fig.~\ref{logic gate}. Let us imagine, conservatively, that during operation half of this structure's total mass ($m = 9 \times 10^{-22} \kilogram$) moves sinusoidally, $p = A \sin (2 \pi f t)$, with an amplitude of $A =10 \, \nanometer$ at a frequency of  $f=100\,\megahertz$. The top speed is then $v_{max} = 2 \pi f A = 6.28 \, \meter / \second$ and the maximum acceleration is $a_{max} = 4 \pi^2 f^2 A = 3.95 \times 10^9 \, \meter / \second^2$. The maximum force applied to the moving mass is then $F_{max} = m \, a_{max} = 3.56 \piconewton$. This force is three orders of magnitude below what is required to break a single carbon-carbon bond ($\approx 6 \nanonewton$). Calculations with HyperChem show that the lateral stiffness of a single molecular joint of the type shown in Fig.~\ref{logic gate} is about $k_{lateral} = 13 \nanonewton / \nanometer$. Lateral deflection of this joint under a load of $3.56 \piconewton$ would be less than a picometer. 

\subsubsection{Effect of Large-Scale Design on Drag}
The friction evaluation of \cite{Hogg_2017} considered complete rotations of a rotor connected to a housing solely through the rotary joints. When used in the mechanical computer discussed in this paper (and as illustrated in Fig.~\ref{logic gate}), the rotor is linked to other parts of the gate.

These links increase coupling to the environment, and could affect drag. For example, the link will reduce the amount of rotor axis tilt and shift, compared to simulations of an isolated rotor and housing. The link, in effect, stiffens the rotor against tilt, and is an alternative to local changes in the design that could also stiffen the joint, e.g., altering the housing size to place the bond in tension rather than compression. These alternatives are an example of how molecular machine design goals could be realized either locally or at larger scales by choice of how the molecular machine is embedded into its environment.

Another instance of a higher-level design choice is arranging designs of neighboring rotary joints to help reduce dissipation. Specifically, the rotary joint has a small potential barrier to rotation~\cite{Hogg_2017} that could be a significant contribution to dissipation at low temperatures. That is, imperfect recovery of energy storage and release when moving over the rotor potential leads to dissipation. This could be reduced by arranging the rotors at each end of a link to be offset by 60 degrees, In this case, when one rotor is at a potential minimum the other is at a maximum. This would keep energy storage and release closer to the rotors than, say, a storage spring located near the clock input, which could be less effective at transmitting energy to and from the rotor due to ``rubbery'' links. This procedure is analogous to using counterweights with elevators.

\section{Conclusions}
Universal combinatorial logic and sequential logic together are known to suffice for the creation of a general purpose, or Turing-complete, computational system. Subject to practical limits of time and memory (as in any computer), such a system can compute anything that can be computed.

We have demonstrated that, using only links and rotary joints, a Turing-complete computational system can be created. Universal combinatorial logic has been demonstrated with the design of a NAND gate, while sequential logic, mimicking electronic flip-flops and sufficient to create memory, has been demonstrated using cells combined into shift registers.

This design approach is far simpler than any other mechanical Turing-complete design of which we are aware. Additionally, due to the avoidance of substantial sliding friction, the ``links and rotary joints" paradigm has the potential to be more power efficient than any previous design of which we are aware. In fact, simulations suggest that molecular-scale implementations of the described system would be far more power efficient than conventional electronic computers.

\section*{Acknowledgments}
The authors thank Damian G. Allis, Jeremy Barton, and Michael S. Marshall for useful feedback on earlier versions of this article.

\appendix
\section*{APPENDIX}
\begin{figure}
\centerline{\includegraphics[width=4.5in]{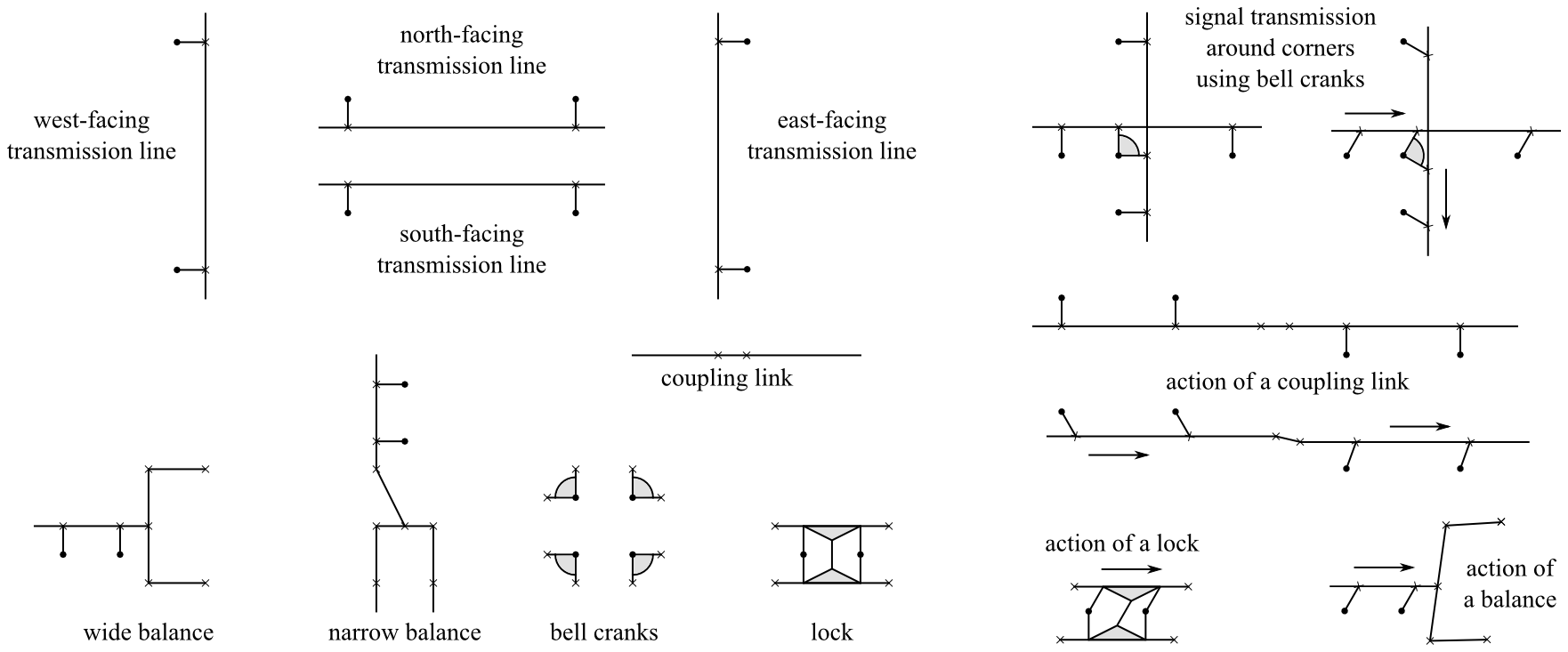}}
\caption{Schematic symbols for mechanical linkage logic components, including the lock, two versions of a balance, and mechanisms for routing signals.}
\label{schematic symbols}
\end{figure}

\begin{figure}
\centerline{\includegraphics[width=4.5in]{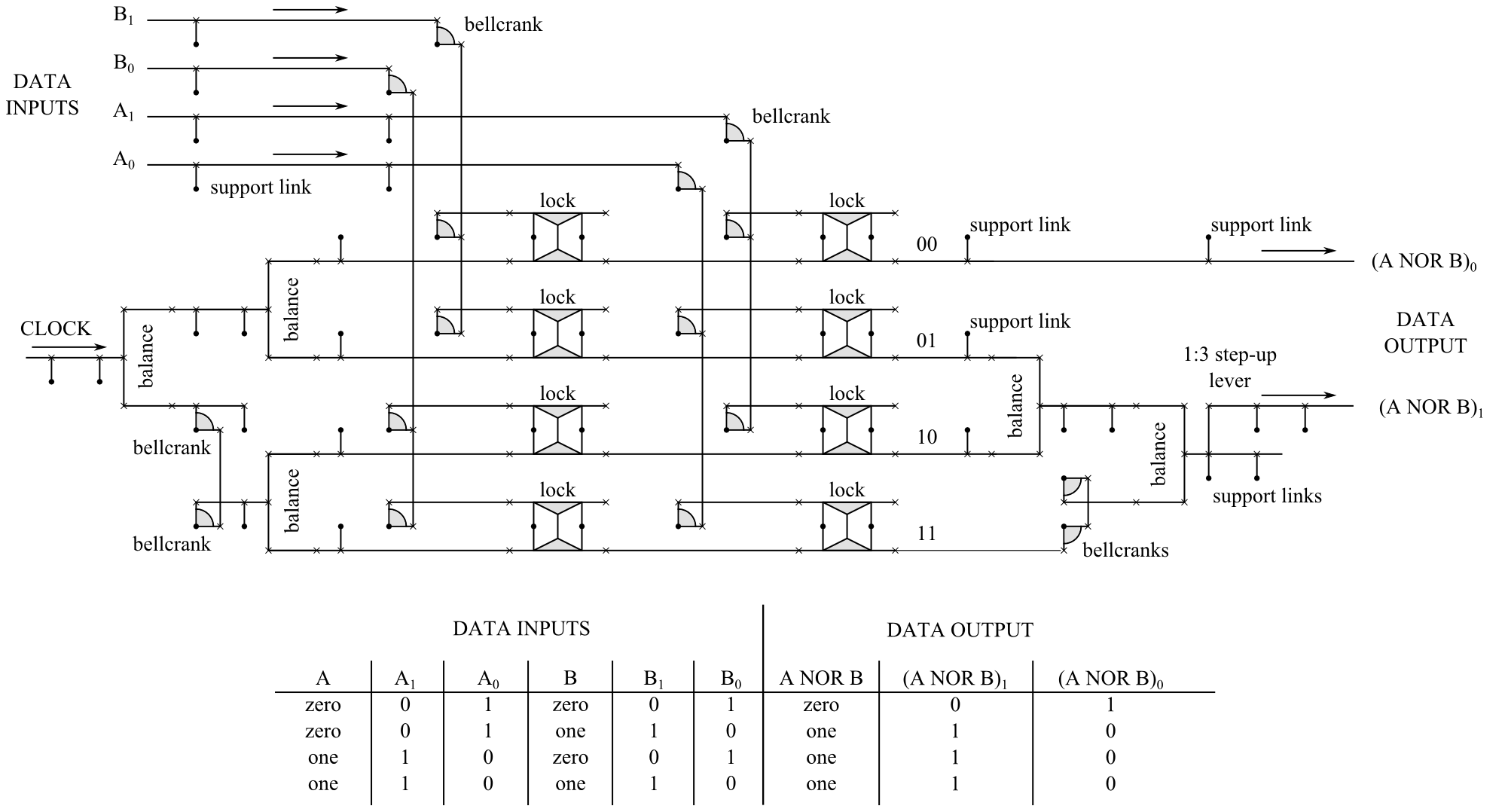}}
\caption{A balance- and lock-based NOR Gate, using a two-link per bit input/output design.}
\label{NOR-gate}
\end{figure}

\begin{figure}
\centerline{\includegraphics[width=4.5in]{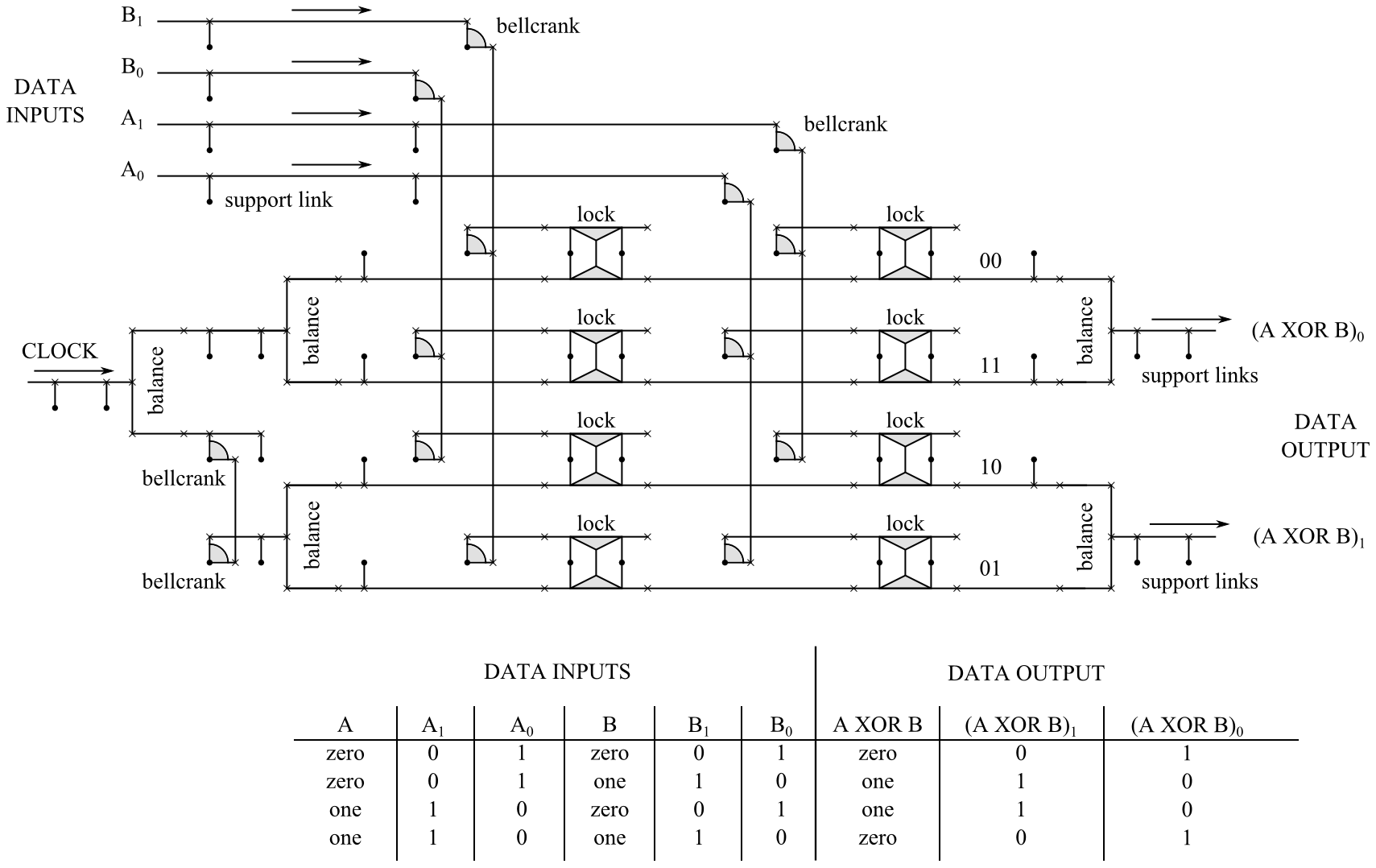}}
\caption{A balance- and lock-based XOR Gate, using a two-link per bit input/output design.}
\label{XOR-gate}
\end{figure}

This appendix contains diagrams that show how common logic gates can be constructed from locks and balances (Fig.~\ref{schematic symbols}). The included gates are: NOR (Fig.~\ref{NOR-gate}) and XOR (Fig.~\ref{XOR-gate}). OR, AND, and XNOR gates can be implemented by inverting the gates shown - inversion is easily accomplished by simply switching the ``one" and ``zero" lines.

\end{document}